\documentclass[3p]{elsarticle}

\usepackage{lineno,hyperref}
\journal{Information Science}

\usepackage{algorithmic}

\usepackage{mathrsfs, booktabs, multirow, makecell}

\usepackage{caption}

\usepackage{url}

\usepackage{color}

\usepackage{subfig}

\usepackage{amsmath,amssymb,amsthm,wasysym,threeparttable,algorithm}
\usepackage{appendix}

\newtheorem{myDef}{Definition}
\newtheorem{myThem}{Theorem}
\newtheorem{corollary}{Corollary}

\newdefinition{rmk}{Remark}
\newproof{pf}{Proof}
\newproof{pot}{Proof of Theorem \ref{thm}}

\begin{document}

\begin{frontmatter}

\title{Privacy-preserving Generative Framework Against Membership Inference Attacks}
% \tnotetext[mytitlenote]{Fully documented templates are available in the elsarticle package on \href{http://www.ctan.org/tex-archive/macros/latex/contrib/elsarticle}{CTAN}.}

% \address{Radarweg 29, Amsterdam}
% \fntext[myfootnote]{Since 1880.}

% %% or include affiliations in footnotes:
% \author[mymainaddress,mysecondaryaddress]{Elsevier Inc}

% \author[mysecondaryaddress]{Global Customer Service\corref{mycorrespondingauthor}}
% \cortext[mycorrespondingauthor]{Corresponding author}

% \address[mymainaddress]{1600 John F Kennedy Boulevard, Philadelphia}
% \address[mysecondaryaddress]{360 Park Avenue South, New York}
\author[1,2]{Ruikang Yang}
\ead{rkyang@stu.xidian.edu.cn}
\author[1,2]{Jianfeng Ma}{}
\ead{jfma@mail.xidian.edu.cn}
\author[1,2]{Yinbin Miao}{}
\ead{ybmiao@xidian.edu.cn}
\author[1,2]{Xindi Ma}{}
\ead{xdma1989@gmail.com}

\address[1]{the State Key Laboratory of Integrated Service Networks (ISN), Xidian University, Xi’an {\rm 710071}, China}
\address[2]{School of Cyber Engineering, Xidian University, Xi'an {\rm 710071}, China}

\begin{abstract}
Artificial intelligence and machine learning have been integrated into all aspects of our lives and the privacy of personal data has attracted more and more attention. Since the generation of the model needs to extract the effective information of the training data, the model has the risk of leaking the privacy of the training data. Membership inference attacks can measure the model leakage of source data to a certain degree. In this paper, we design a privacy-preserving generative framework against membership inference attacks, through the information extraction and data generation capabilities of the generative model variational autoencoder (VAE) to generate synthetic data that meets the needs of differential privacy. Instead of adding noise to the model output or tampering with the training process of the target model, we directly process the original data. We first map the source data to the latent space through the VAE model to get the latent code, then perform noise process satisfying metric privacy on the latent code, and finally use the VAE model to reconstruct the synthetic data. Our experimental evaluation demonstrates that the machine learning model trained with newly generated synthetic data can effectively resist membership inference attacks and still maintain high utility.
\end{abstract}


\end{frontmatter}

% \linenumbers

\section{Introduction}
With the widespread use of artificial intelligence and machine learning in our daily life, the need to ensure privacy protection in these systems has come to the fore, with an overarching goal that the output and decision do not leak specific secrets of true individual attributes. Nowadays, machine learning has been provided as a service (MLaaS) by many platforms\footnote{https://cloud.google.com/automl}\footnote{ https://azure.microsoft.com/en-us/services/machine-learning}\footnote{https://aws.amazon.com/cn/}. Smart devices access these models through prediction Application Programming Interfaces (APIs) which return a prediction score vector and some companies are willing to deploy these models in smart devices locally. Meanwhile, several architectures have been proposed \cite{chilimbi2014project} \cite{mcmahan2017communication} for distributed and federated learning to protect privacy of the clients’ training data. Each client trains a local model on his own data and periodically exchanges model parameters or partially constructed models with the other clients. However, a series of studies \cite{shokri2017membership, nasr2019comprehensive, salem2018ml, song2019privacy, nasr2018machine, jia2019memguard, song2021systematic, hui2021practical, zhang2021membership, choquette2021label} have indicated that the adversary could exploit the prediction output or parameters of machine learning models to perform membership inference attacks to get helpful information about the data used in the machine learning model. Membership inference attack \cite{shokri2017membership}, as one of the most important and concerning attack modes, is always used to measure the leakage of the machine learning model \cite{nasr2019comprehensive}. Given an input record and a target model, the membership inference attack tries to determine whether the record was used during the training of the target model or not.

Membership inference attacks cause severe privacy and security threats, especially when the training dataset contains sensitive attributes such as diagnosis and income. For example, if a hacker could determine a patient data is used to train a cancer prediction model which deployed in a wearable device, he was able to judge that the patient have cancer with great probability. As a consequence, this patient’s privacy would be disclosed.

In machine learning, the influence of a user on trained model should be erased at request as the learned model contains private information about its training data. These follow the spirit of the European Union’s General Data Protection Regulation (GDPR) \cite{gdpr-web}. Meanwhile, the main techniques to achieve the privacy preservation in machine learning are secure multi-party computing (MPC) and fully homomorphic encryption (FHE). However, the MPC scheme applied to protect the security of machine learning cannot effectively resist similar inference attacks because it doesn't protect the final result, and the FHE scheme is not widely used in practical application environments because of its complex calculations. Therefore, how to design a practical and effective mechanism to resist the membership inference attack is still an unsolved challenge in machine learning. 

Overfitting is the major cause for the feasibility of membership inference attack, as the trained model tends to memorize training inputs and perform better on them. As a result, the prediction scores or the parameters' gradients in the target model are distinguishable for members and non-members of the training data. There are several defenses \cite{shokri2017membership, nasr2019comprehensive, salem2018ml, song2019privacy, nasr2018machine, song2021systematic, wang2020against, yin2021defending} have been proposed and most of them try to reducing overfitting by various regularization technique, such as L2 regularization \cite{shokri2017membership}, min-max game based adversarial regularization \cite{nasr2018machine, yin2021defending}, and dropout \cite{salem2018ml}. In addition, Song  \emph{et al.}\cite{song2021systematic} showed that early stopping outperformed other overfitting-prevention counter-measures against membership inference attacks to classifiers. However, none of them directly reduce the distinguishability of members or non-members and make the big utility loss. Some other mechanisms \cite{abadi2016deep, bassily2014private, wang2017differentially, mcmahan2018general, yu2019differentially, iyengar2019towards, jayaraman2019evaluating} like Differentially Private - Stochastic Gradient Descent (DP-SGD) leverage differential privacy \cite{dwork2006calibrating} when training the target model. In addition, some research like MemGuard\cite{jia2019memguard} and Purification framework\cite{yang2020defending} leverage both the specific differential privacy mechanism and machine learning constraint method to the prediction output of target model. However, all of these work consider that leverage some mechanisms in the training stage or the output of the target model, but actually, data and model reuse is common in practice, and each time we utilize these data to train a new model or apply these machine learning models, we have to adjust the settings of the mechanism for the corresponding situation. Therefore, in this paper, we consider that some transformation mechanisms could apply to the input data to prevent membership inference attacks fundamentally. 

% I wish you the best of success.

 

In this paper, we propose a privacy-preserving modular generative framework to defend membership inference attacks by transforming the source data into synthetic data to train the target model. In the general situation where the same set of data is used for training different models, the privacy protection mechanism tampering with the training phase of the target model need change according to the different models and the model convergence faces challenges. Meanwhile, the noise mechanism applied to the output will also reduce the accuracy of the target model to a certain extent. While some data generation mechanisms\cite{xie2018differentially, chen2020gs, zhang2018differentially, torkzadehmahani2019dp} satisfying differential privacy have been proposed, all of these works consider that leverage differential noise mechanism in the training stage as previously stated and the membership inference attacks are not included. The biggest problem of these work is that they should consider more about the model convergence and privacy budget costs in each iteration, which makes the model hard to converge and extend. On the other hand, their purpose is to construct generative models that satisfy differential privacy to generate lots of new data, while this paper focuses on protecting the target model from membership inference attacks by transforming the original data into similar data that satisfy differential privacy.

\textbf{The goal of this study is to develop a data generation method for machine learning that provides a provable privacy against membership inference attacks, without compromising much utility.}

In this paper, the membership inference attack, including black-box attack and white-box attack, is considered a measurement to reflect the privacy leakage of the trained model about its training set.

We achieve the generative framework by generating synthetic data through a well-trained Variational Autoencoder (VAE) generative model on source data. {Unlike many other schemes\cite{hayes2019logan, chen2019gan, chen2021gan, liu2019performing, hu2021membership} that set the goal of protection as a generative model, we treat VAE as a building component to protect the downstream task model, which is the difference between our work among them.}
We treat the VAE model as a specific data generation solution because the VAE model creates a continuous and complete latent space locally and the new never seen data could be generated. 
Compared with the traditional generation model that randomly generates hidden vectors in the hidden space, we generate new data through the hidden vector generated from the existing original data, which can maintain certain distribution characteristics of the data to be generated. And our mechanism can generate new data that is more in line with the target data distribution when facing the scene of partial data generation. Notice that we choose to generate new data rather than stay the latent vectors because of the portability of downstream training tasks.

In conclusion, we make the following contributions in the paper.

\begin{itemize}
  \item  We propose a privacy-preserving modular generative framework to generate robust representation of the source data for all type of machine learning and this framework can be further easily fine-tuned for many different downstream tasks such as classification as shown in our work. 
  \item  We adopt VAE as a specific extractor and reconstructor in our framework, because of the stochasticity of the generative model and the continuity of the VAE model, which is unlike the aforementioned others set the generative model as the protection target. Moreover, we adopt the metric privacy mechanism in the latent space of VAE rather than the data domain. Note that, we don't apply the differential privacy mechanism in the model during the training process like many papers mentioned above.
  \item  We evaluate our system framework against the state-of-art membership inference attack on three real-world datasets. Our results show that our project can protect data resources without much efficiency loss.
 
\end{itemize}

In the remainder of this article, we firstly describe preliminaries in Section \ref{section2}. Then, we present the system model and threat model in Section \ref{section3}. After that, Section \ref{section4} proposes our privacy-preserving generative framework and the related algorithms. Experiments and results are discussed in Section \ref{section5} and related works are shown in Section \ref{section6}. Conclusions of this article with future work are presented in Section \ref{section7}.

\section{Preliminaries}\label{section2}
In this section, we review three concepts used in our work, namely, generative model, differential privacy, and membership inference attack. We briefly introduce the variational autoencoder used as transformation function and the metric privacy which is applied in satisfying privacy guarantee.
\subsection{Generative Model}

\textbf{Variational Autoencoder\cite{kingma2013auto} (VAE):} VAE is a widely used generative framework that consists of an encoder and a decoder, which are cascaded to reconstruct data with pre-defined similarity metrics. The encoder maps data into a latent space, and the decoder maps the encoded latent code back to the data space. The VAE objective function is composed of the reconstruction loss and the prior regularization over the latent code distribution. Formally, 

\begin{equation}
\min\limits_{\theta,\phi}-\mathbb{E}_{q_{\phi}(z|x)}[p_{\theta}(x|z)] + KL(q_{\phi}(z|x)||P_z)
\end{equation}

where $z$ denote the latent code, $x$ denotes the input data, $q_{\phi}(z|x)$ is the probabilistic encoder parameterized by $\phi$ which is used to approximate the true posterior, $p_{\theta}(x|z)$ represents the probabilistic decoder parameterized by $\theta$, and $KL(\cdot||\cdot)$ denotes the KL divergence. In practice, $q_{\phi}(z|x)$ is always constrained to be Gaussian distribution and is sampled via the reparameterization trick, which results in a closed-form derivation of the second term. 

\subsection{Differential Privacy}

Differential privacy \cite{dwork2006calibrating} is one of the most popular paradigm used to deal with information leak in statistical databases. It provides a formal privacy guarantee, ensuring that sensitive information relative to individuals cannot be easily inferred by disclosing answers to aggregate queries.

A mechanism $K$ is $\epsilon$-differential private if for any two adjacent databases $x$, $x'$, and any property $Z$, the probability distributions $K(x)$, $K(x')$ differ on $Z$ at most by $e^{\epsilon}$, which means \begin{math}K(x)(Z) \leq e^{\epsilon}K(x')(z)\end{math}. 

\subsubsection{Metric Privacy}
While the standard differential privacy is a rigorous privacy notion, it is only applicable to publishing aggregate statistics. The problem studied in this paper wishes to generate data, rather than aggregate statistics about the data. Therefore, it requires a more general privacy notion for data that belongs to an arbitrary domain of secrets. The authors of \cite{chatzikokolakis2013broadening} extended the principle of differential privacy and proposed a generalized notion, metric privacy. Essentially, it defines a distance metric between secrets and guarantees a level of indistinguishability proportional to the distance. Specifically, given an arbitrary set of secrets $\mathcal{X}$ with a metric $d_{\mathcal{X}}$:

\begin{myDef}[\textbf{Generalized Privacy \cite{chatzikokolakis2013broadening}}]
  \label{dx-privacy} A mechanism from $\mathcal{X}$ to $\mathcal{Z}$ is a (probabilistic) function $K$:$\mathcal{X}$$\rightarrow$$P(\mathcal{Z})$ which satisfies $d_{\mathcal{X}}$-privacy, iff $\forall x,x'\in\mathcal{X}:d_{\mathcal{P}}(K(x),K(x'))\leq d_{\mathcal{X}}(x,x')$, or equivalently:
  \begin{equation}
    K(x)(Z) \leq e^{d_{\mathcal{X}}(x,x')}K(x')(Z) \qquad \forall Z \in \mathcal{F}_{\mathcal{Z}}
  \end{equation}
  where $\mathcal{Z}$ is a set of query outcomes, $\mathcal{F}_{\mathcal{Z}}$ is a $\sigma$-algebra over $\mathcal{Z}$, and $P(\mathcal{Z})$ is the set of probability measures over $\mathcal{Z}$.
\end{myDef}

Metric privacy guarantees that the output of a mechanism should be roughly the same, i.e., bounded by the distance $d_{\mathcal{X}}(x,x')$, between two inputs $x$ and $x'$. For an adversary who observes the output space, it is challenging to infer the exact input, thus the privacy of the input is protected. 
The metric $d_{\mathcal{X}}$ can be extended by scaling a standard metric $d$ by a factor $\epsilon$, i.e., $d_{\mathcal{X}}=\epsilon \cdot d$. As a result, the guarantee of metric privacy also relies on $\epsilon$: lower $\epsilon$ indicates higher indistinguishability, hence stronger privacy.

\subsection{Membership Inference Attack}
We formulate the membership inference attack as a binary classification task where the attacker aims to classify whether a sample has been used to train the victim machine learning model. Formally, we define 
\begin{equation}
\mathcal{A}:(x,\mathcal{M}(\theta)) \rightarrow \{0,1\}
\end{equation}
where the attack model $\mathcal{A}$ output 1 if the attacker infers that the sample $x$ is included in the training set, and 0 otherwise. $\theta$ denotes the target model parameters while $\mathcal{M}$ represents the general model publishing mechanism, i.e., type of access available to the attacker. For example, $\mathcal{M}$ is an identity function for the white-box access case and can be the inference function for the black-box case.

\section{System Model and Threat Model}\label{section3}
In this section, we first describe the system model of our framework and then elaborate on the threat model defined in our framework. 

\begin{figure}[!t]
\centering
\includegraphics[width=3in]{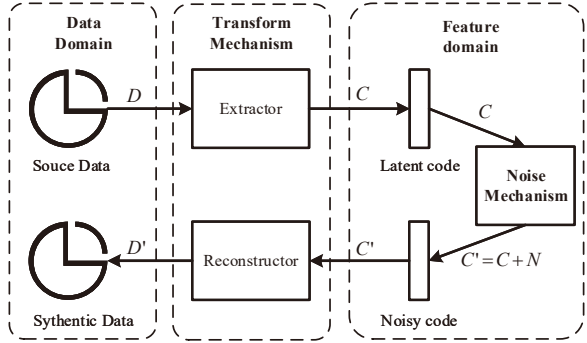}
% where an .eps filename suffix will be assumed under latex, 
% and a .pdf suffix will be assumed for pdflatex; or what has been declared
% via \DeclareGraphicsExtensions.

% \captionsetup{justification=centering}
\caption{Privacy-preserving generative framework.}
\label{fig_1}
\end{figure}

\subsection{System Model}
We depict in \figurename \ref{fig_1} the proposed system model for a privacy-preserving generative framework to train the target machine learning model against membership inference attacks. 
At the very outset, the data owner builds an extractor and a reconstructor through their data $D$. Then he adds noise $N$ which satisfy differential privacy to the output $C$ of the extractor and gain synthetic data $D'$ from reconstructor. At the same time, if the source data has label originally, the label of synthetic data should be created. In the end, synthetic data could be used to train new models. 

More specifically, the data owner first trains a fundamental VAE model $M_{G}$ locally (see \ref{sec:Extractor and Reconstructor}) using the data source $D$. 
The VAE model maps the original data to the latent space vector through the encoder and then maps the latent space vector back to the original data through the decoder. Because the decoder can rebuild the data in original data through the latent space vector, it shows that the latent space vector contains enough original data information.
Some noise (see \ref{sec:noise addition mechanism}) $N$ is added to the latent vector generated by model $M_{G}$ and the data owner leverages the decoder of $M_{G}$ to reconstruct data $D'$. If necessary, implement label migration through label-transfer mechanism (see \ref{sec:Label-Transfer Mechanism}). Then we could use this synthetic set $D'$ to train an application model $M_{A}$.

\subsection{Threat Model}
\noindent
\textbf{Attacker's goal:} 
Like many studies on membership inference attacks, we consider an attacker's goal is to determine whether an input sample $x$ was used as part of the training set $D'$. But in our solution, we focus on protecting the raw dataset $D$ rather than the synthetic dataset $D'$. Thus, we replace $D'$ with $D$. 
In short, in this paper, we focus on the privacy of the source data $D$. \\
\textbf{Attacker’s capability:} 
Because there are three important entities in our solution, we need to classify attackers (shown in TABLE \ref{MIA}) based on the knowledge of framework. Firstly, we assume that they only have the ability to access the application model after data generated. With the knowledge of model structure and parameters, we can separate them into black-box attack and white-box attack. These two types of attacks are almost the same as the traditional membership inference attack mode: black-box attack and white-box attack except the protected data. In the other situation, the attacker can get the synthetic data. Inspired by the Chen \cite{chen2019gan}, according to the knowledge of the extractor and reconstructor, the attack mode can be divided into a black-box generator attack, partial black-box generator attack, and white-box generator attack. These three types of attacks belong to the attack category facing data release and the generative model release. \textbf{Note that, in this paper, we just consider the following black-box attack and white-box attack}.

\begin{table}[htbp]
\renewcommand{\arraystretch}{1}
% if using array.sty, it might be a good idea to tweak the value of
% \extrarowheight as needed to properly center the text within the cells
\caption{Taxonomy of Membership Inference Attacks}
\label{MIA}
\centering
\tabcolsep 16pt
% Some packages, such as MDW tools, offer better commands for making tables
% than the plain LaTeX2e tabular which is used here.
% \resizebox{\textwidth}{5mm}{
% \begin{tabular}{lcccc}
\begin{tabular}{lp{1.4cm}<{\centering}p{1.4cm}<{\centering}p{1.4cm}<{\centering}p{1.4cm}<{\centering}}
% \begin{tabular}{lcccc}
\toprule
Name  & extractor & reconstructor & synthetic data & application model \\
  \midrule
  \textbf{black-box} & \CIRCLE & \CIRCLE & \CIRCLE & $\times$ \\
  \textbf{white-box} & \CIRCLE & \CIRCLE & \CIRCLE & \checkmark  \\
  black-box generator & \CIRCLE & \CIRCLE & \Circle & \checkmark  \\
  partial black-box generator & \RIGHTcircle & \RIGHTcircle  & \Circle & \checkmark  \\
  white-box generator & \Circle & \Circle  & \Circle & \checkmark  \\
  \bottomrule
\end{tabular}
% }
\begin{tablenotes}
\item[1] $\CIRCLE$:without access; $\RIGHTcircle$: with access; $\Circle$: full control.
\item[2] $\checkmark$: white-box; $\times$: black-box.
\end{tablenotes}
\end{table}

\noindent\textbf{Taxonomy of membership inference attacks discussed in this paper:}
\begin{enumerate}
\item \textbf{black-box attack:} This is the least knowledgeable setting to attackers where the attacker is only able to send query data samples to the target application model and obtain their output predicted by the target model. Besides, they have no information about the extractor and reconstructor in this framework.
\item \textbf{white-box attack:} The attacker has access to the full application model, including its architecture and parameters, and any hyper-parameter that is needed to predict in the model. At the same time, he can also observe the intermediate computations at hidden layers. However, he also doesn't know the extractor and reconstructor in this framework.
\end{enumerate}

\section{Privacy Preserving Generative Framework}\label{section4}

This section details the implementation of our privacy preserving generative framework (PPGF) to enforce both privacy and validity. Our method focuses on preserving privacy and our idea is inspired by the mechanism in \cite{fan2019practical}. The source data will first be extracted to be feature vectors. 
Next, add noise into the feature vectors to achieve privacy guarantees and go through the reconstruction function resulting in the synthetic data. 

In \cite{fan2019practical}, Fan \emph{et al.}, utilize the SVD to decompose the image matrix and add noise to fuzzy the perceptual similarity of each image. But we want to generate new data that satisfies the distribution of the source data, so the reconstructor needs to have generalized properties. This means that giving suitable input to the reconstructor, we could generate new data that is never seen but satisfy the distribution of source data in some features.

In the following sections, we first discuss the overview of our framework and then elaborate on the three main components. Finally, the privacy analysis  of PPGF is described formally.

\begin{figure*}[!t]
\centering
  \subfloat[Initial Stage with Label Generation Stage]
  {
    \includegraphics[width=2.5in]{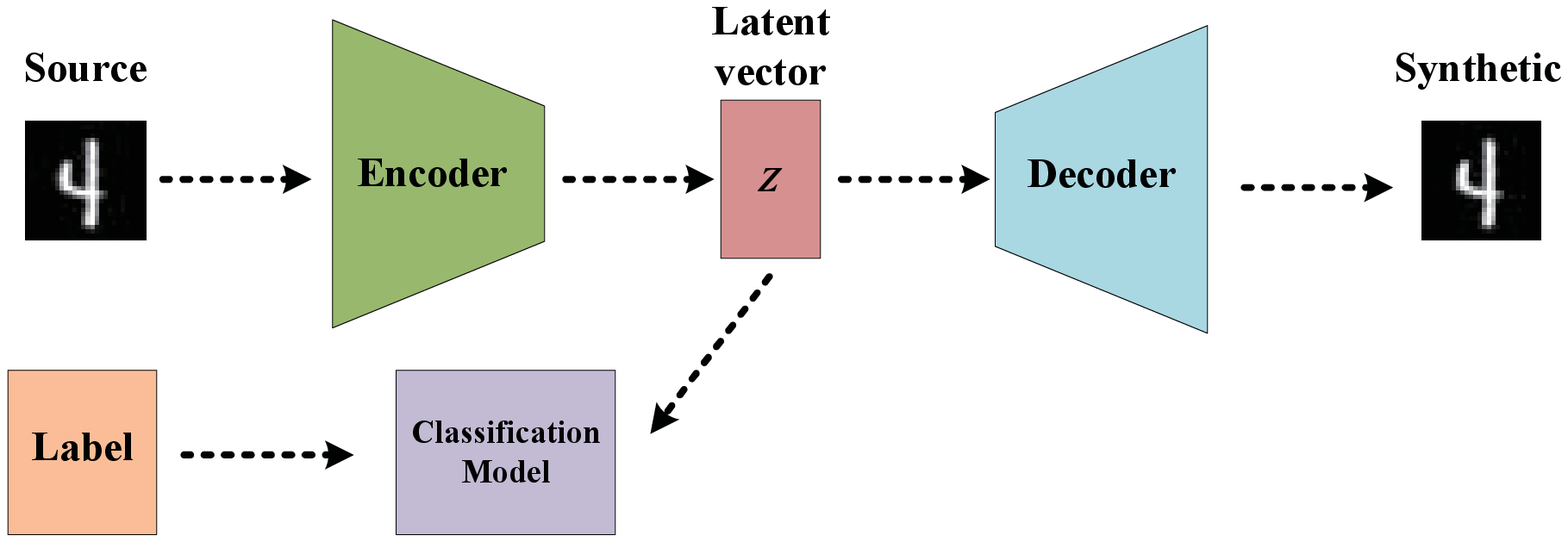}
    \label{overview1}
  }
  \subfloat[Data Generation Stage with Label Generation Stage]
  {
    \includegraphics[width=2.5in]{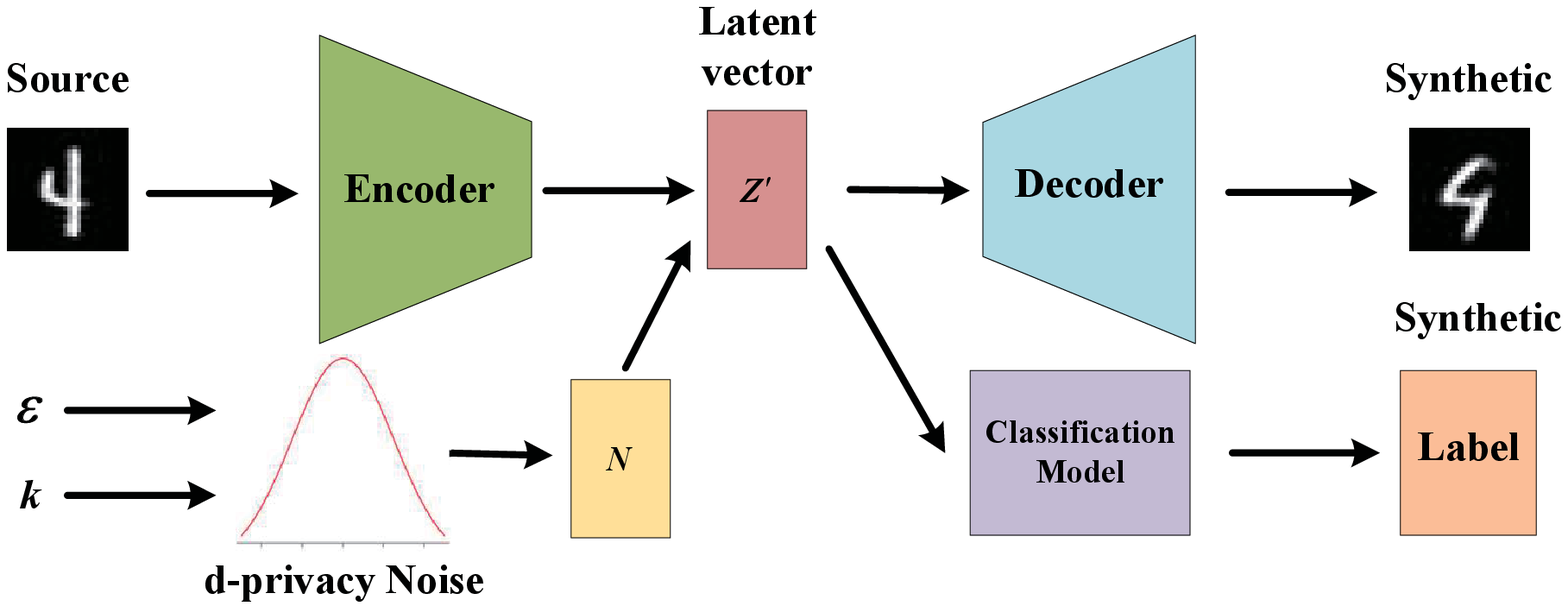}
    \label{overview2}
  }
% where an .eps filename suffix will be assumed under latex, 
% and a .pdf suffix will be assumed for pdflatex; or what has been declared
% via \DeclareGraphicsExtensions.

% \captionsetup{justification=centering}

\caption{Overview of privacy-preserving system framework.}
\label{fig_2}
\end{figure*}

\subsection{Overview of the PPGF}
\begin{figure}[!t]
\centering
\includegraphics[width=2.5in]{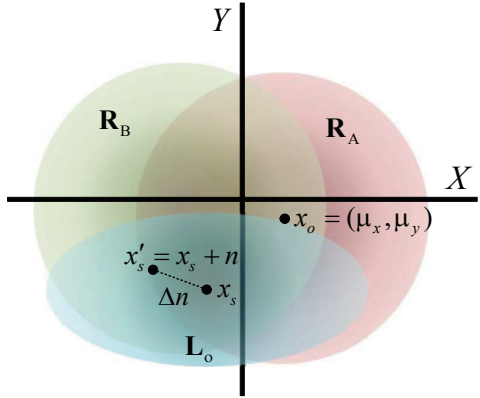}\\
% where an .eps filename suffix will be assumed under latex, 
% and a .pdf suffix will be assumed for pdflatex; or what has been declared
% via \DeclareGraphicsExtensions.

% \captionsetup{justification=centering}

\caption{Procedure in latent space.}
\label{fig_3}
\end{figure}

 As shown in \figurename \ref{fig_2}, our basic scheme can be divided into two main phases and an optional phase. The initial stage and the data generation stage are the main stages, and the label generation stage is an optional stage. In addition, all of these stages are executed locally. 
\begin{enumerate}
\item In the initial stage, our main task is to select or generate suitable \textbf{extractor and reconstructor}. In this solution, the VAE model is used as the extractor and reconstructor. As shown in \figurename \ref{fig_2}\subref{overview1}, by training a generator that can generate original data, map the data to the latent space to get $Z$ specified by the encoder in the VAE model and the decoder maps the latent space data $Z$ back to the data space. At this time, we treat encoder as the mapping relationship $A$ from the source data $D$ to the latent space and treat decoder as $A^{-1}$, which means $A$ is reversible.

\item In the data generation stage, our main task is to use the existing mapping relationship $A$ and a given privacy budget $\epsilon$ to generate synthetic data that meets the differential privacy requirements and can resist membership inference attacks. In this scheme, we leverage a \textbf{noise addition mechanism} that satisfies $d-privacy$. As shown in \figurename \ref{fig_2}\subref{overview2}, we first map the source data $D$ into the latent space through the mapping relationship $A: A(D) \rightarrow Z$, and use the noise mechanism $M$ to disturb the latent vector $M(Z) \rightarrow Z' = Z + N$, and finally, the mapping relationship $A^{-1}$ is used to map $Z'$ back to the source data space to obtain $D' \leftarrow A^{-1}(Z')$. For instance, as shown in \figurename \ref{fig_3}, we assume that our latent space is a two-dimensional space. We firstly process an original data point $x_1$ through the VAE encoder to obtain the point $x_o$ in latent space and the corresponding variance $\theta_o$, where the data point $x_1$ is mapped to get the red area $R_A$ that satisfies $N(x_o,\theta_o)$ distribution in latent space. Then we randomly select a point in the area $R_A$ to get $x_s$. After that, the noise mechanism obtains the corresponding blue area $L_o$ that satisfies the probability density function according to the area $R_A$, and we randomly selects ${x_s}'$ in the noise area $L_o$. Then VAE decoder decodes ${x_s}'$ into data $x'$. Note that, $R_B$ is the green area that another data $x_2$ is mapped in the latent space. 

\item In the label generation stage, we mainly provide labels for the newly generated data through the \textbf{label-transfer mechanism} if the original data contains labels. Through the existing mapping relationship $A$, as shown in Fig. 2(a), we put the source data $D$ into the latent space: $A(D) \rightarrow Z$. Then we use the data in the latent space and the corresponding labels $(Z, L)$ to train a classification model {$M_c$} in the latent space. Next, as shown in Fig. 2(b), we use $Z'$ in latent space generated in the data generation stage and the label classification model {$M_c$}, and we can get the label $L'$ of the new data $D'$.
\end{enumerate}

The components of our basic scheme are mainly included: (1) extractor and reconstructor, (2) noise addition mechanism, and (3) label-transfer mechanism. We will separately describe these three components in detail in the next sections.

\subsection{Extractor and Reconstructor} \label{sec:Extractor and Reconstructor}
We first describe the components, Extractor and Reconstructor, in detail of our scheme. In general, the extractor and reconstructor appear in pairs. After the extractor gains the features from the input, the paired reconstructor could recover the data from these features as much as possible. In this paper, we consider the type of input data as images. We adopt the VAE or VAE-structure generative model to extract data features and reconstruct new data because VAEs learn encoders that produce probability distributions over the latent space instead of points in the latent space and are easy to converge. In addition, when we sample from these probability distributions during many training iterations, we effectively show the decoder that the entire area around the distribution’s mean produces output that are similar to the input value. In short, we create a continuous and complete latent space locally and we could generate the new never seen data. However, we need to point out that the limitation of the VAE model. We need to treat the prior probability as the Gaussian form and Gaussian distribution is a bad representation of low-dimensional data. At the same time, if an adversary can access the VAE model, the VAE model is vulnerable under the membership inference attack (shown in \cite{hilprecht2019monte}).

% VAE machine learning model is also this mode. 
There are an encoder and a decoder in the VAE machine learning model, and we consider the latent layer's output as the feature information. We treat the VAE model as a combination of extractor and reconstructor. Each time, even we train a VAE model using the same data set and initial hyper parameters, the parameters of the final generated model is different. So, we could treat the generated VAE model as a stochastic extractor and stochastic reconstructor. 

 We denote the extractor as \begin{math}EC:\mathbb{R}^{*} \rightarrow \mathbb{R}^{k}\end{math}, which maps an input data $D$ to a k-dimensional real vector $z \in \mathbb{R}^{k}$. We denote VAE model with $EC_{\rm vae,k}$, which means we consider the VAE encoder as the extractor, and the dimension of the latent layer is $k$. As for reconstructor, it is like an invertible function of extractor in this solution. 
 
 If the feature vector is created by the paired extractor, the output of the reconstructor is more similar to the input of the extractor. We denote the reconstructor as \begin{math}DC:\mathbb{R}^{k} \rightarrow \mathbb{R}^{*}\end{math}, which maps a k-dimensional real vector $z$ to a synthetic data. And we denote the VAE model with $DC_{\rm vae,k}$, which means we consider the VAE decoder as the reconstructor, and the dimension of the latent layer is k. Therefore, we put the mapping relationship $A$ as $EC_{\rm vae,k}$ and put the inverse mapping $A^{-1}$ as $DC_{\rm vae,k}$.

\subsection{Noise Addition Mechanism} \label{sec:noise addition mechanism}
In this section, we describe the procedure of adding noise in detail. Given the input feature vector $x_{0}$, the private mechanism $K$ performs random in space $\mathbb{R}^{k}$ according to certain probability distributions which can provide differential privacy for $x_{0}$. The following theorem states the privacy guarantee and differential privacy can be generalized to the case of an arbitrary set of secrets $\mathcal{X}$, equipped with a metric $d_{\mathcal{X}}$ and privacy budget $\epsilon$. Note that in the framework of $\epsilon d$-privacy, we can express the privacy of the mechanism itself, on its own domain, without the need to consider a query. In our scenario, using the Hamming metric of standard differential privacy, which aims at completely protecting the value of an individual, would be too strong. We are not interested in completely hiding the data characteristics, since some approximate information needs to be revealed to obtain the required service. Hence, using a privacy level that depends on the Euclidean distance between the latent vectors is a natural choice. Privacy proof of the Theorem \ref{thm1} is described in the paper\cite{chatzikokolakis2013broadening}. 

\begin{myThem} \label{thm1} In a k-dimensional space, a mechanism $K$ that samples $x$ according to the following probability density function satisfies $\epsilon d_k$-privacy
\begin{equation}
\label{equation_dprivacy}
D_{\epsilon,k}(x_0)(x)=\lambda_{\epsilon,k}e^{-\epsilon\cdot d_k(x_0)(x)}
\end{equation}
% for the sake of simplicity, 
where $d_k(x_0)(x)$ represents k-dimensional Euclidean distance\footnote{Euclidean distance can measure similarity of images in the eyes of humans and encoder can preserve some image perceptual similarity} and k is without loss of generality.
\begin{equation}
\lambda_{\epsilon,k}=\frac{1}{2}\cdot(\frac{\epsilon}{\sqrt{\pi}})^k\cdot\frac{(\frac{k}{2}-1)!}{(k-1)!}
\end{equation}
The coefficient $\lambda_{\epsilon,k}$ is derived in Appendix A.
\end{myThem}

Generating $x$ according to (\ref{equation_dprivacy}) can be achieved as follows in Algorithm \ref{algorithm1}. We first set $x$ at the origin of the hyper-spherical coordinate system. Then we sample the radial coordinate $r$ according to the marginal distribution and uniformly sampling a point on the unit $(k-1)$-sphere. We multiply the results of them and add them with $x$ to gain the new feature vector $x'$. Note that, the marginal distribution satisfies the $\Gamma$-distribution and this proof is shown in Appendix B. 

\begin{algorithm}
  \caption{Noise Addition Mechanism}
  \label{algorithm1}
  \begin{algorithmic}[1]
  \REQUIRE A given feature vector $x \in \mathbb{R}^k$; a given privacy budget $\epsilon$.
  \ENSURE A new $d$-privacy feature vector $x'$
  \FOR {$i$$\leftarrow$ 1,2,3,...,k}
    \STATE Random sample $r_i$ from Gaussian distribution ${\rm \mathcal{N}(0,1)}$
  \ENDFOR
  \STATE Compute $\hat{r}=\frac{1}{|(r_1, r_2, ...,r_k)|} \cdot (r_1, r_2, ...,r_k)$
  \STATE Random sample $p$ from uniform distribution ${{U}\rm {(0,1)}}$.
  % \STATE Random select $d$ from pdf $f(d , \epsilon, k)$ of $\Gamma$-distribution.
  \STATE Set $d = C_{\epsilon, k}^{-1}(p)$, where $C_{\epsilon, k}^{-1}(p)$ is the inverse function of cumulative distribution function (cdf) of $Gamma(k, \epsilon)$.
  \STATE Compute $x' = x + \hat{r} \cdot d $ 
  \RETURN $x'$
  \end{algorithmic}
\end{algorithm}

\begin{myDef}
  \label{sensitivity} {\rm (\textbf{$l_2$-sensitivity})}: $l_2$-sensitivity $\delta$ of function $f$ indicates the maximum distance between dataset $\mathcal{D}$ and $\mathcal{D'}$
  \begin{equation}
    \delta = max||f(\mathcal{D})-f(\mathcal{D'})||_2
  \end{equation}
\end{myDef}

For the VAE model, we set the input feature vector $z$ as the value obtained by sampling from $\rm \mathcal{N}(\mu, \sigma^2)$ according to the output of Encoder $EC$. Thus, the $k$ is set to be the dimension of $z$. As for the $\epsilon$, because each dimension in the hidden layer is approximated to a standard Gaussian distribution in the well-trained VAE model, we use the empirical rule \cite{kazmier1976schaum} to estimate the max distance, namely sensitivity according to Definine \ref{sensitivity} where $f(x)=x$. We set $\epsilon$ as $\frac{\epsilon}{(3\sigma)}$ where $\epsilon \in (0,1)$ and $\sigma$ is generated by Encoder $EC$.

\subsection{Label-Transfer Mechanism} \label{sec:Label-Transfer Mechanism}
Sometimes, for semi-supervised or supervised learning models, some source data have labels at the beginning. However, through our generative framework, while the raw data is transformed into new data, the labels of the new data may not correspond one-to-one with the source data. Therefore, during semi-supervised or supervised learning, for labeled source data, we need to use label-transfer algorithms to label new data. In this paper, we adopted a model-driven label-transfer mechanism and the specific algorithm in Algorithm \ref{algorithm2}.

\begin{algorithm}
  \caption{Model-Driven Label Transfer Mechanism}
  \label{algorithm2}
  \begin{algorithmic}[1]
    \REQUIRE Chosen extractor $EC$; source data $D$ and its label $L_D$; the latent layer vectors of target data $v_T$; classifier model $M_c$
    \ENSURE Label of target data $L_T$
    \FOR {$i$$\leftarrow$ 0,1,2,...}
      \STATE $v_{D_{i}} \gets EC(D_i)$
      \STATE Add $v_{D_{i}}$ to Training Data Set $D_T$
    \ENDFOR
    \STATE Train $M_c$ through data set $D_T$ and label $L_D$
    \STATE Predict $L_T \leftarrow M_{c}(v_T)$
    \RETURN $L_T$
  \end{algorithmic}
\end{algorithm}

We first use the feature extractor $EC$ in the framework to extract the feature vectors of the source data set $D$ with labels $L_D$ and then train a classification model $M_c$ based on the corresponding labels $L_D$ and these feature vectors $Z_T$. After that, put the previously recorded feature vector $Z_T'$ used in the newly generated data into this classification model for prediction, and then the label $L_T$ corresponding to the newly generated data can be obtained. Note that, this classification model $M_c$ is sensitive facing a white-box generator attack.

For the VAE model, we use our label-transfer mechanism after the VAE model is well trained. If we gain the well-trained VAE model, the label-transfer mechanism records the feature vector $z$ generated by the Encoder according to the source data $d \in D$ and the corresponding tag $l \in L$. When the data generative mechanism is running, we record the noised feature vector $z'$ and the corresponding generated data $d'$. The classification model $M_c$ leverages the feature vector $z$ and label $l$ obtained from the source data set for training, and the well-trained classification model $M_c$ predicts the label $l'$ of the corresponding data $d'$ according to the noised feature vector $z'$.

\subsection{Privacy Analysis of PPGF}

In this section, we provide a concise privacy analysis of our framework. In our framework, there are two main stages and we only need to make sure the output of the data generation stage satisfying $\epsilon d_k$-privacy privacy. We make the claims that the synthetic data satisfies $\epsilon d_k$-privacy privacy. Without loss of generality, we mark the encoder as the mapping relationship $A$ and the decoder as the mapping relationship $A^{-1}$. We process the raw data $x$ through the encoder $EC$ to get $A(x)$ and utilize Algorithm 1 to add noise $n$ to get new latent code $A(x)+n$, at this time $A(x)+n$ satisfies $\epsilon d_k$-privacy according to the Theorem 1. Finally, the decoder $DC$ of the VAE model is used to obtain the synthetic data $x'=A^{-1}(A(x)+n)$. 

We first give the Theorem \ref{thm2} to show that applying differential privacy mechanism $M$ to the data in the latent layer can also make the decoded data meet the same privacy requirements.

\begin{myThem} \label{thm2} {\rm (\textbf{Post-Processing \cite{dwork2014algorithmic}}):} Let $M$ be an $(\epsilon, \delta)$-differential private function and let $F:R \rightarrow R'$ be an arbitrary mapping where $R'$ is any arbitrary space. Then $F \circ M$ is $(\epsilon, \delta)$-differential private.
% \lable{thm2}
\end{myThem}

The above properties based on $(\epsilon, \delta)$-differential privacy can also be extended to $\epsilon d_k$-privacy, because $\epsilon d_k$-privacy is an extension of $(\epsilon, \delta)$-differential privacy. The following Corollary can be obtained.

\begin{corollary} \label{col1} {\rm (\textbf{Extended Post-Processing}):} Let $M$ be an $\epsilon d_k$-privacy probability density function and let $F:R \rightarrow R'$ be an arbitrary mapping where $R'$ is any arbitrary space. Then $F \circ M$ is also $\epsilon d_k$-privacy.
\end{corollary}

According to Corollary \ref{col1}, we know that applying $\epsilon d_k$-privacy mechanism $M$ to latent vector can also make the decoded data meet the same privacy requirements and we get Theorem \ref{thm3}. 

\begin{myThem} \label{thm3} The synthetic data $x'=A^{-1}(A(x)+n)$ satisfies $\epsilon d_k$-privacy.
\end{myThem}

\section{Experiments}\label{section5}
Based on the existing membership inference attacks, we present the most comprehensive evaluation to the proposed privacy-preserving generative framework. Our experiments include verification of the validity and utility of the synthetic data according to the accuracy of the training task, and the analysis of the defense of our framework according to the some metric of the membership inference attack. Moreover, we used three common datasets and utilized two efficient membership inference attack methods.

There are three baselines in our experiments: (1) BS (Source data), (2) N-PPGF (Synthetic data with non-private), and (3) DP-SGD. BS represents training each model based on source data set. N-PPGF where the noise addition mechanism is removed from PPGF is designed to show the upper bound on expression ability of the VAE model used in the PPGF. DP-SGD is designed according to the work of \cite{abadi2016deep} \cite{mcmahan2018general} which is the generic privacy protection method and widely used. {We should note that although we usually treat $\epsilon$ as the measure of privacy as we called privacy budget in this paper, we could calculate $\epsilon$ from $\sigma$ in DP-SGD. Therefore, we use $\sigma$ to measure the privacy of DP-SGD. The $\delta$ is privacy relaxation parameter\cite{mcmahan2018general} for DP-SGD and we set $\delta=1e-05$ for DP-SGD all the time in this paper.}
% The experiments result can show the advantages of the proposed generative framework against membership inference attacks and the availability of the newly generated data is acceptable. 
% At the same time, corresponding verification was made for the principle that the scheme can resist membership inference attacks. 

We implemented our framework in Pytorch. We trained all of the models on a PC equipped with one Titan V GPU with 12 GB of memory. We used Scipy and Numpy for samping from $\Gamma$-distribution. The parameters were set to default values according to the databases, i.e., $\epsilon = 0.5$ and $k = 20$ for MNIST. We used Opacus\footnote{https://opacus.ai/} to implement DP-SGD method. 

\subsection{Setup}

This part is mainly the setup phase of the experiment, including the databases, models, and measurement for various tests.

\subsubsection{Databases}
As stated in TABLE \ref{database}, There are three databases MINST\footnote{http://yann.lecun.com/exdb/mnist/}, Fashion-MNIST\footnote{https://github.com/zalandoresearch/fashion-mnist} and CelebA\footnote{http://mmlab.ie.cuhk.edu.hk/projects/CelebA.html} used in our experiments and all of them are widely used in Image learning field. 

\begin{table}[htbp]
\renewcommand{\arraystretch}{1}
% if using array.sty, it might be a good idea to tweak the value of
% \extrarowheight as needed to properly center the text within the cells
\caption{Databases}
\label{database}
\centering
\tabcolsep 4pt
% Some packages, such as MDW tools, offer better commands for making tables
% than the plain LaTeX2e tabular which is used here.

\begin{tabular}{lcccc}
\toprule
  Name&Category&Pixel Size&Set Size \\
  \midrule
  MNIST &Handwritten digits&28$\times$28&70,000\\
  Fashion-MNIST&Zalando's article&28$\times$28 & 70,000  \\
  CelebA &Face attributes&178 $\times$ 218 & 202,599   \\
  \bottomrule
\end{tabular}

\end{table}

\subsubsection{VAE models}
We choose two different VAE generative model for these three datasets. The details of the architecture of the VAE model for MNIST and FashionMNIST is presented in TABLE \ref{Architecture of VAE model for MNIST and FashionMNIST} and the details of the architecture of the VAE model for CelebA is shown in TABLE \ref{Architecture of VAE model for CelebA} in Appendix C. Note that, we crop the image size in CelebA database from 178 $\times$ 218 to 128 $\times$ 128 when training the VAE model. And we resize the generated image from 128 $\times$ 128 to 178 $\times$ 218 when training the target model.

\subsubsection{Target models}
We choose target models with varying capacity, including a small convolution model as CNN used for the MNIST, deep ResNet\cite{he2016deep}, and SlimCNN\cite{sharma2019slim} used for CelebA. The small convolution model contains 2 convolution layers with kernels 32 and 64, a global pooling layer and two fully connected layer of size 9216 and 128. The small model is trained for 20 epochs with initial learning rate 0.2. The small convolution model CNN, ResNet-18,  and ResNet-34 are for MNIST image classification. The FashionMNIST image classification models are ResNet-18, ResNet-34, and ResNet-50. The SlimCNN image classification is specially for CelebA. The detailed configurations and training recipes for deep ResNets and SlimCNN can be found in the original paper. Note that we only use these models for performance testing due to the limitations of the GPU memory size. For security testing of membership inference attacks, we uniformly use the ResNet-18 model, and in particular, we use ResNet-18 on the CelebA dataset to train for gender classification.

\subsubsection{Metrics}
There are two basic evaluation indicators to measure the validity of the synthetic data and the effectiveness of defense mechanism.
\begin{itemize}
  \item \textbf{Prediction accuracy:}
  In order to reflect the validity of the newly generated image data, we separately use these generated data and source data to train the same training task, and compare the prediction accuracy of the different naturally trained model. Besides, the aforementioned N-PPGF baseline and DP-SGD baseline are used in the same experimental conditions to show that the newly generated data can retain somewhat the domain characteristics of the original dataset.

  \item \textbf{Attack accuracy:} 
  {The output of the membership inference attack is divided into two categories: member and non-member. The attack accuracy is used to indicate the source of unknown input data for successful prediction. In order to show the effectiveness of defense, we separately use these generated data and source data to train the same target training task for membership inference attacks. Besides, we calculate the attack prediction accuracy, including precision, recall, fscore. At the same time, to show the classification ability of the attack classifier more accurately, we also use the Area Under Curve (AUC) to accurately express the ability of membership inference attack.} 

\end{itemize}

\subsection{Validity of PPGF}
We generate new data ${D}$ through our framework on the three datasets of MNIST, Fashion-MNIST and CelebA respectively. And through the following experiments to illustrate the validity of the generated data for training tasks.

\begin{table*}[htbp]
\renewcommand{\arraystretch}{1}
% if using array.sty, it might be a good idea to tweak the value of
% \extrarowheight as needed to properly center the text within the cells
\caption{validity of different database on several models}
\label{validity of the generated data}
\centering
\tabcolsep 2pt
\begin{tabular}{|l|c|c|c|c|c|c|c|c|}
\hline 

\multirow{2}{*}{} &

\multicolumn{3}{c|}{MNIST} &

\multicolumn{3}{c|}{FashionMNIST} &

\multicolumn{1}{c|}{CelebA}\\

\cline{1-8}

\textbf{Architecture} & \textbf{CNN} & \textbf{ResNet18} & \textbf{ResNet34} & \textbf{ResNet18} & \textbf{ResNet34} & \textbf{ResNet50} & \textbf{SlimCNN\cite{sharma2019slim}}\\
 % \\

\hline
TA (PPGF $\epsilon=0.5$) & 0.9807 & 0.9938 & 0.9919 & 0.8905 & 0.8685 & 0.8909 & 0.8531\\
VA (PPGF $\epsilon=0.5$) & 0.9741 & 0.9828 & 0.9788 & 0.7823 & 0.7820 & 0.7736 & 0.8016\\
\hline
TA (N-PPGF) & 0.9818 & 0.9838 & 0.9914 & 0.8932 & 0.8897 & 0.8913 & 0.8839\\
VA (N-PPGF) & 0.9835 & 0.9818 & 0.9857 & 0.8138 & 0.8308 & 0.7737 & 0.8011\\
\hline
TA (BS) & 0.9973 & 0.9955 & 0.9969 & 0.9798 & 0.9757 & 0.9709 & 0.9189\\
VA (BS) & 0.9919 & 0.9843 & 0.9918 & 0.9115 & 0.9152 & 0.9023 & 0.9117\\
\hline
TA (DP-SGD $\sigma=0.7$) & 0.8042 & 0.9274 & 0.9413 & 0.7766 & 0.7955 & 0.7686 & 0.8067\\
VA (DP-SGD $\sigma=0.7$) & 0.9291 & 0.9365 & 0.9451 & 0.7716 & 0.7850 & 0.7639 & 0.8092\\
\hline

\end{tabular}
\begin{tablenotes}
\item[1]   TA: Train Accuracy; VA: Validity Accuracy.
\end{tablenotes}
\end{table*}

\subsubsection{Validity of the Generated Data}

\begin{figure*}[htbp]
\centering
  \subfloat[CNN]
  {
    \includegraphics[width=2.0in]{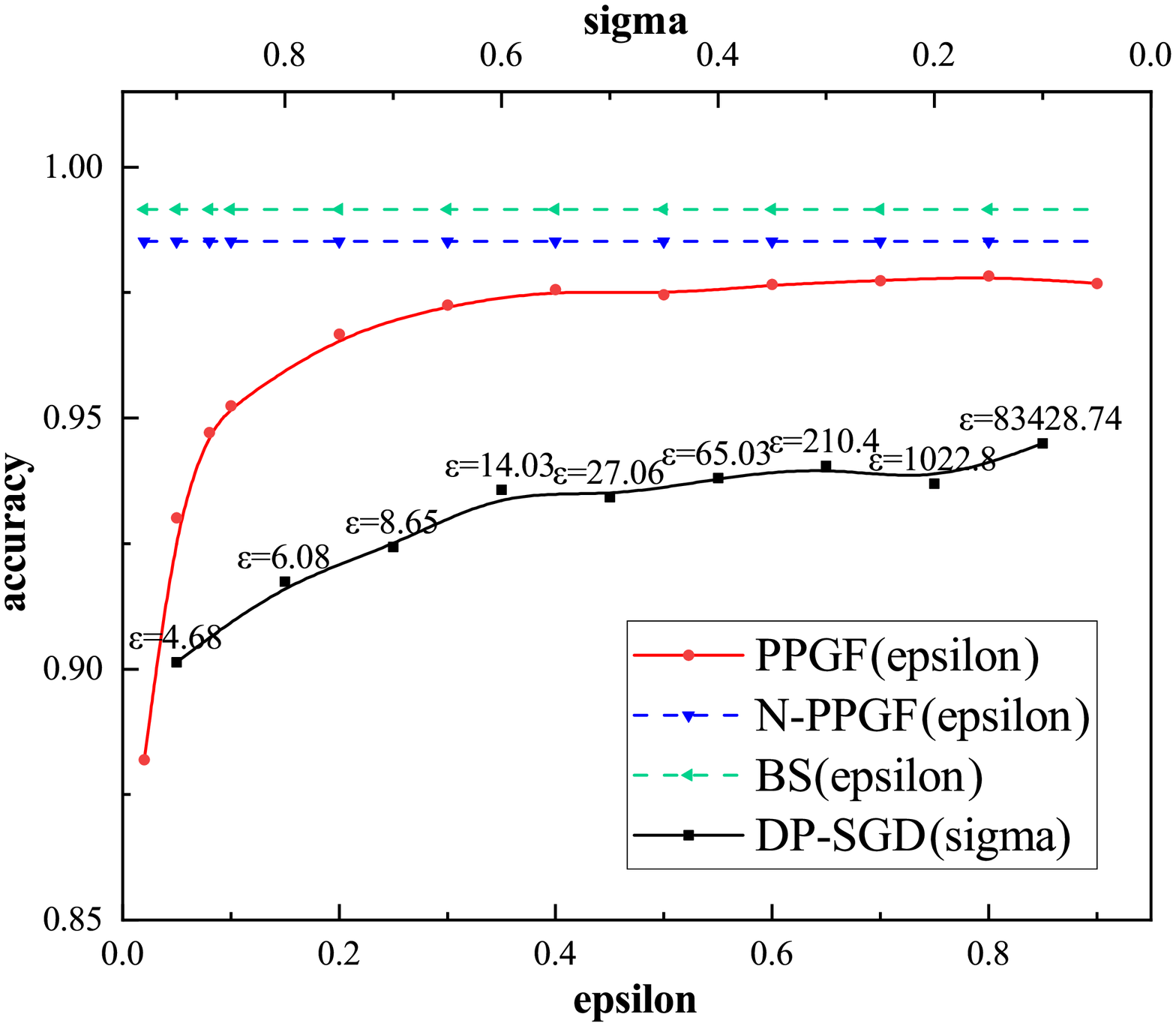}
    \label{M1}
  }
  \subfloat[ResNet$18$]
  {
    \includegraphics[width=2.0in]{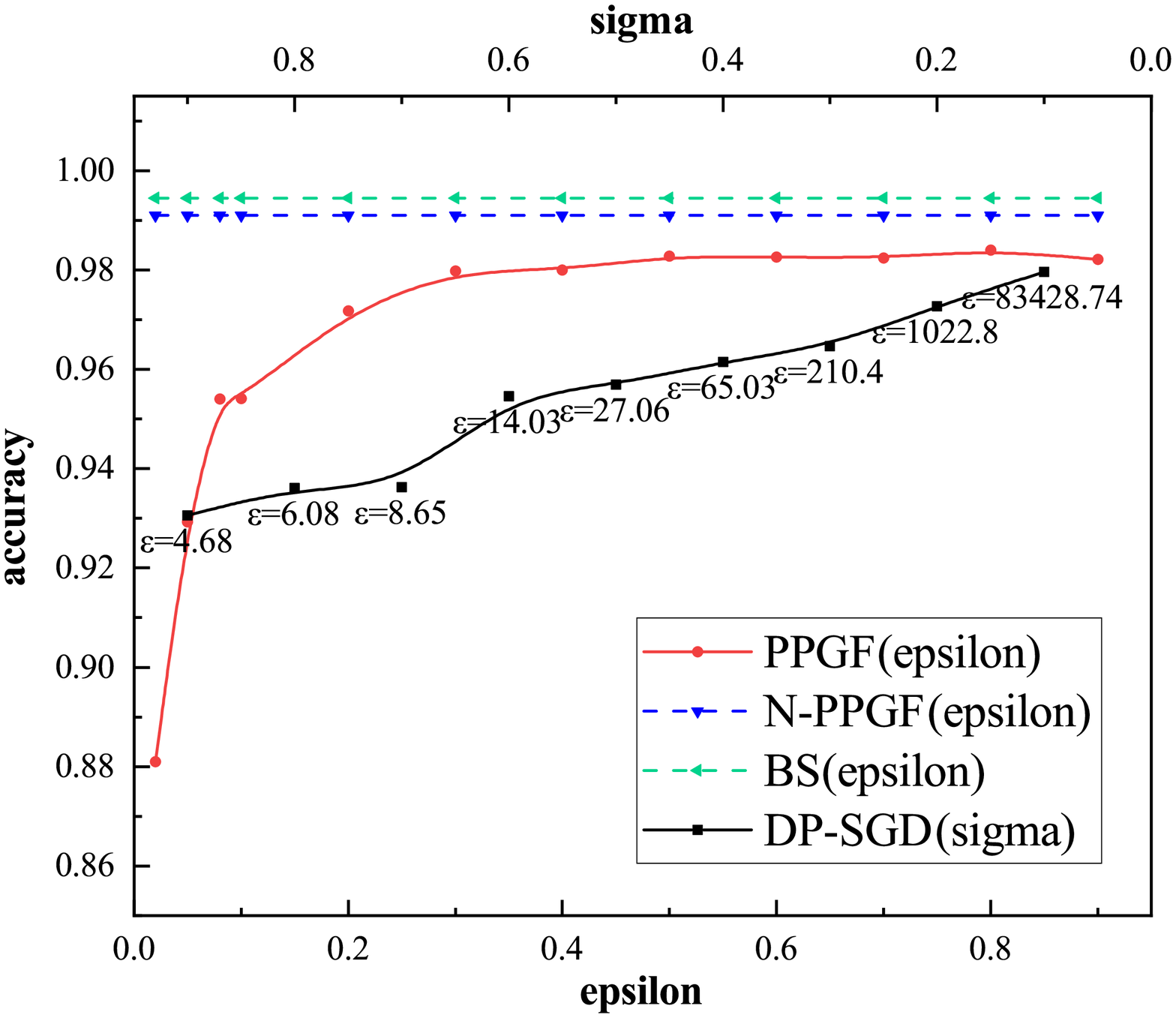}
    \label{M2}
  }
  \subfloat[ResNet$34$]
  {
    \includegraphics[width=2.0in]{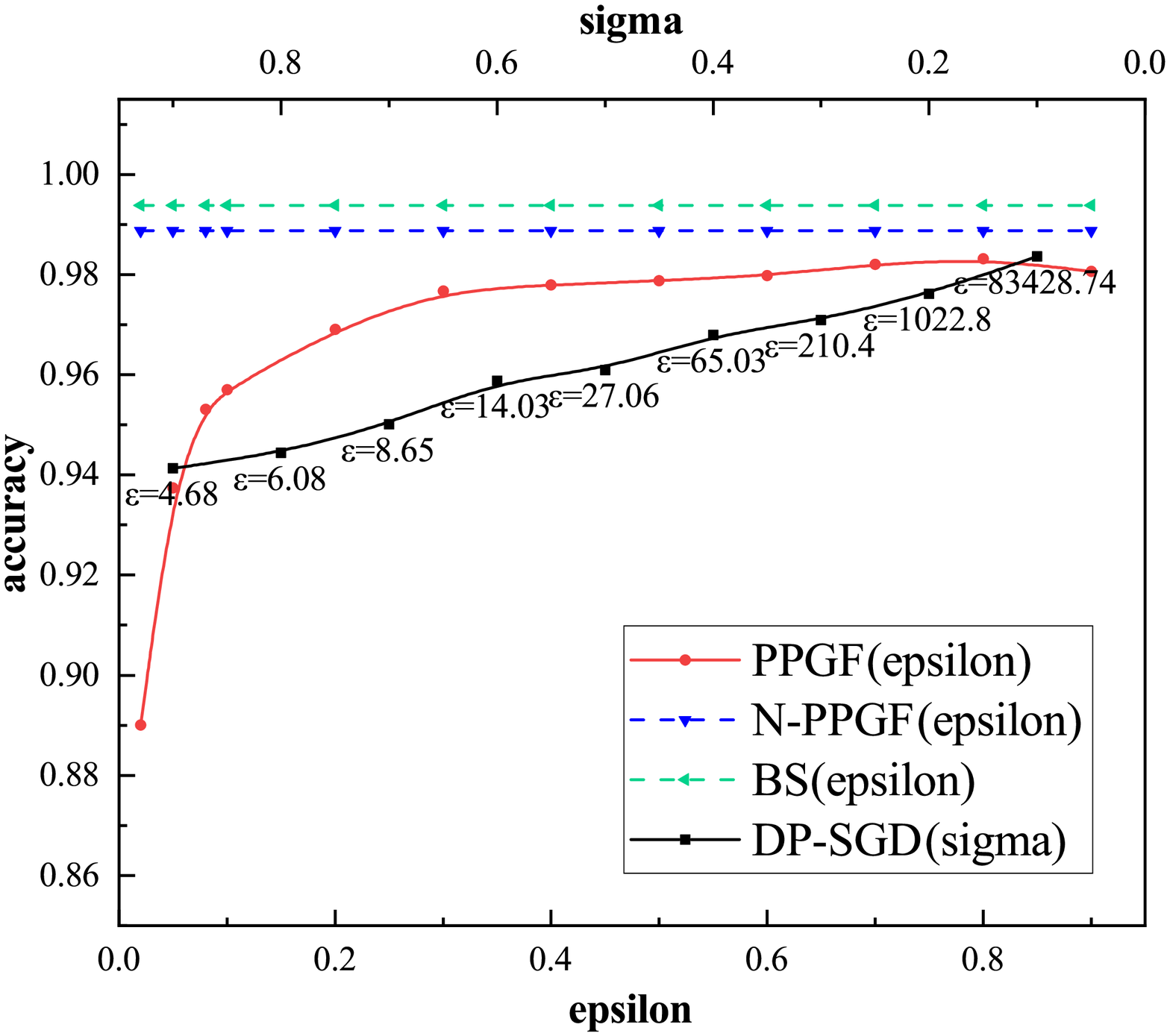}
    \label{M3}
  }
% \captionsetup{justification=centering}
\caption{Prediction Accuracy of MNIST}
\label{fig_4}
\end{figure*}

In order to measure the validity of the generated data, we separately train the same classification training task using source data and the newly generated data. We firstly separately use total training data of these three databases to train the VAE generative model and generate corresponding synthetic data for all training data of these databases. The original dataset are tested on these models under three baselines, and the results are shown in TABLE \ref{validity of the generated data}. It can be seen that through our framework, under the same classification task and the same model, the newly generated dataset with $\epsilon=0.5$ has little decrease in the accuracy of prediction compared with DP-SGD mechanism ($\sigma=0.7$, $\epsilon=8.65$).

\begin{figure*}[htbp]
\centering
  \subfloat[ResNet$18$]
  {
    \includegraphics[width=2.0in]{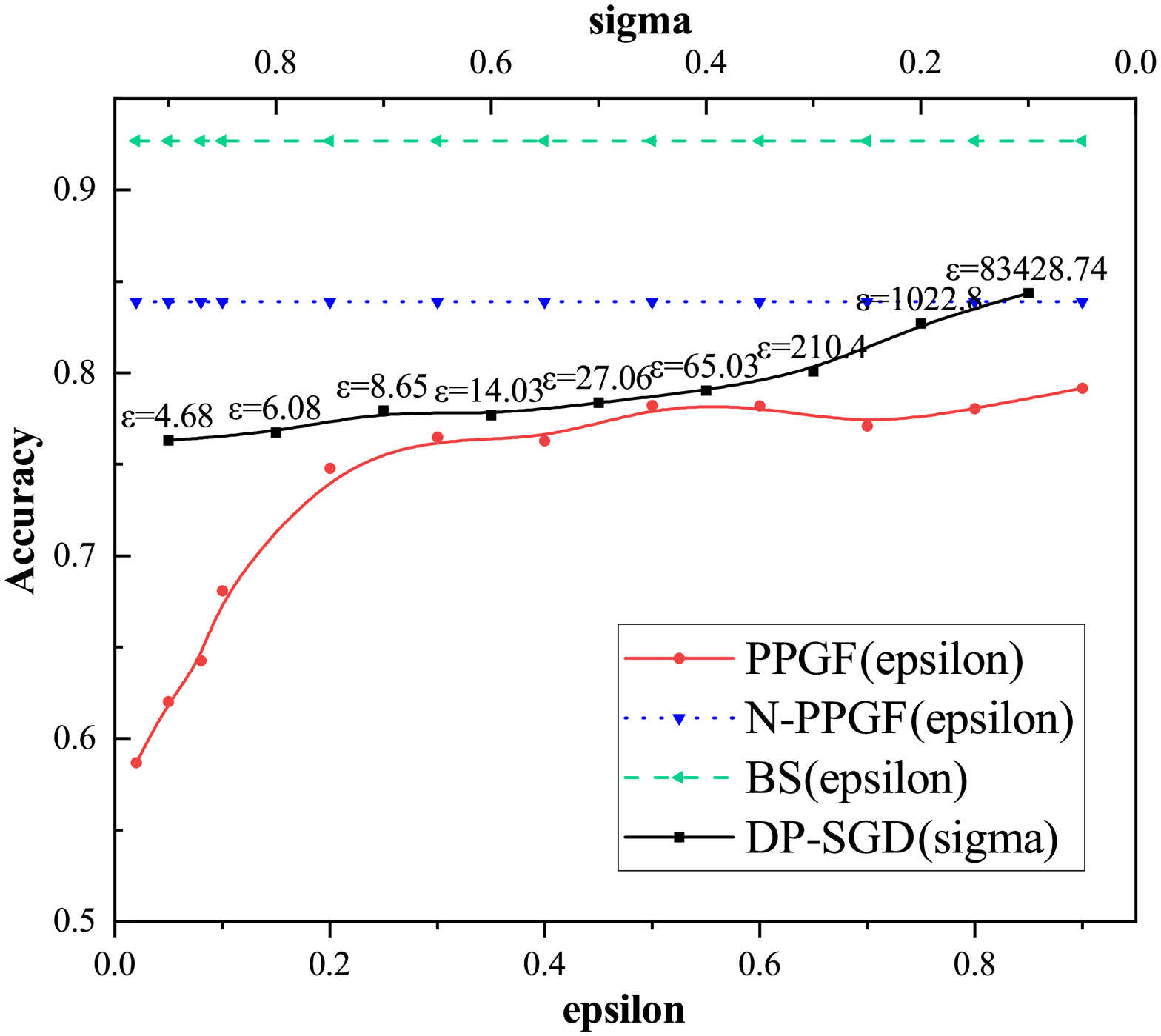}
    \label{M4}
  }
  \subfloat[ResNet$34$]
  {
    \includegraphics[width=2.0in]{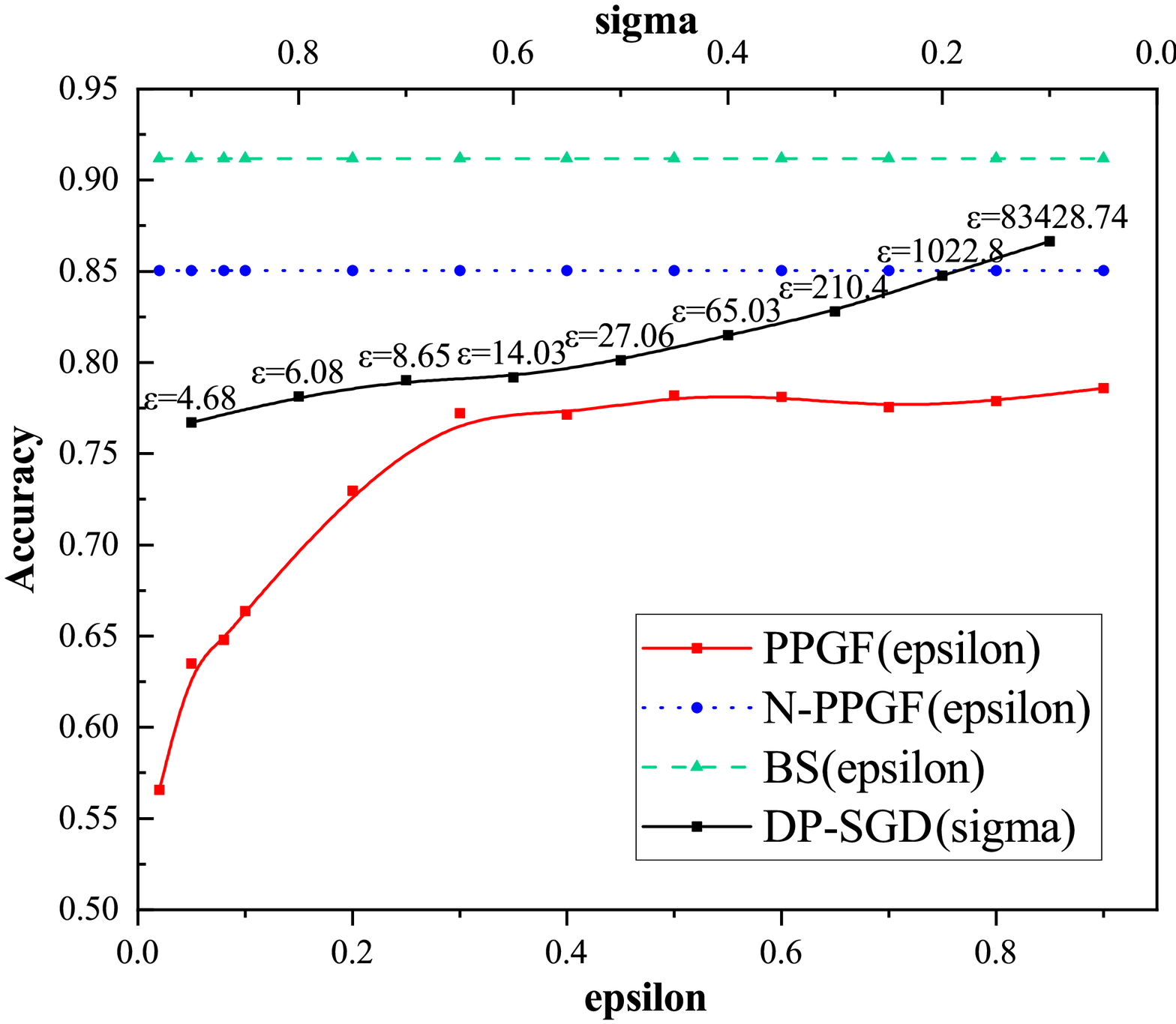}
    \label{M5}
  }
  \subfloat[ResNet$50$]
  {
    \includegraphics[width=2.0in]{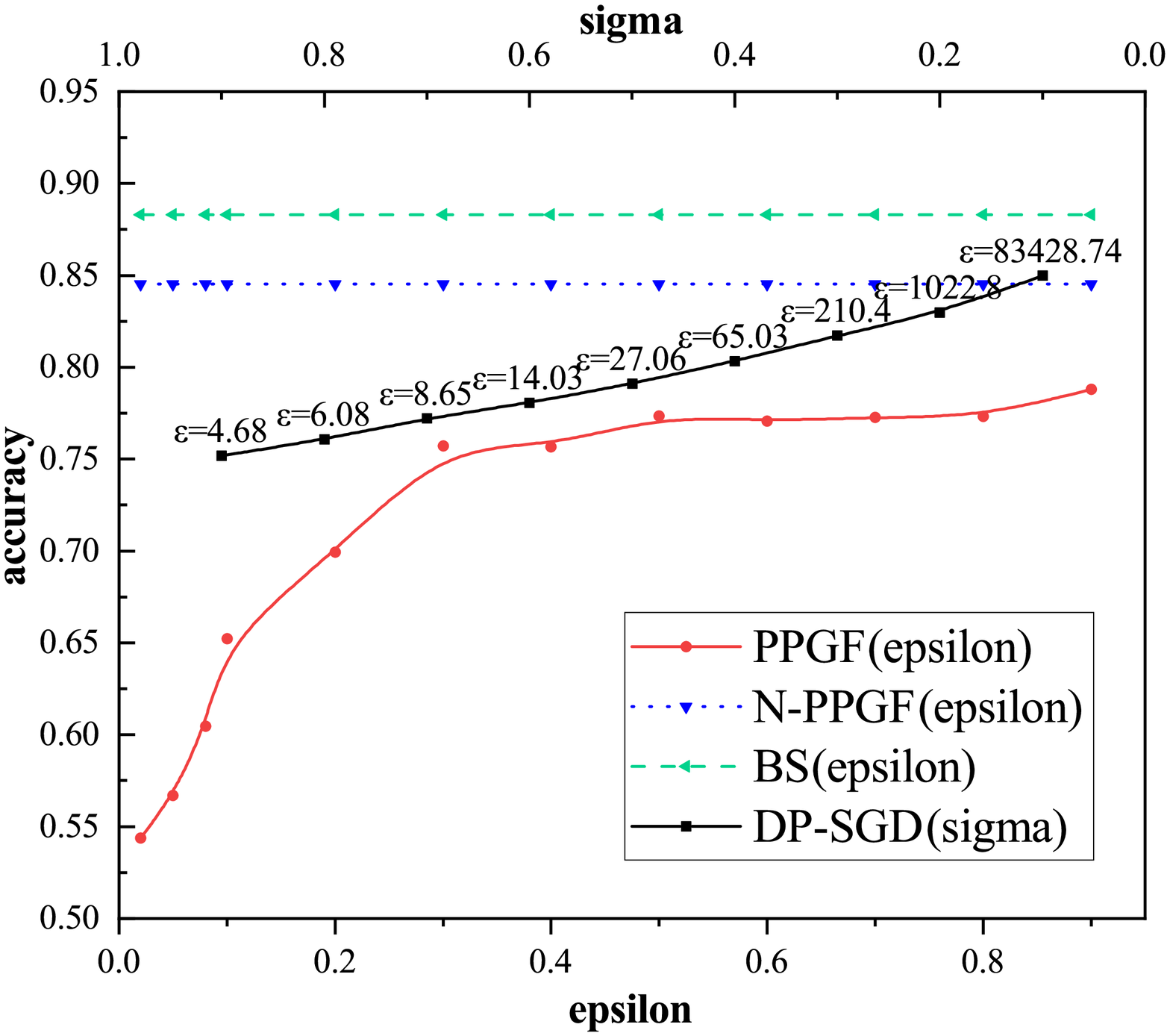}
    \label{M6}
  }
% where an .eps filename suffix will be assumed under latex, 
% and a .pdf suffix will be assumed for pdflatex; or what has been declared
% via \DeclareGraphicsExtensions.

% \captionsetup{justification=centering}

\caption{Prediction Accuracy of FashionMNIST}
\label{fig_5}
\end{figure*}

We also present the privacy utility tradeoff by varying the parameter $\epsilon$ in the following range [0.02, 0.05, 0.08, 0.1, 0.2, 0.3, 0.4, 0.5, 0.6, 0.7, 0.8, 0.9]. Recall that $\epsilon$ can be used to tune the level of privacy as in noise addition mechanism. Lower $\epsilon$ indicates higher indistinguishability, stronger privacy and vice versa. As for DP-SGD, we set $\delta=1e-05$ and $\alpha$ in the following range [1.1, 1.2 ,..., 11, 12, ..., 64]. The parameter $\sigma$ is vary from 0.1 to 0.9. We plot the prediction accuracy of different model for MNIST and FashionMNIST in \figurename \ref{fig_4} and \figurename \ref{fig_5}. The BS baseline, N-PPGF baseline and DP-SGD baseline are included for reference. As can be seen in \figurename \ref{fig_4}, when increasing $\epsilon$, from more private to less private, the accuracy of model starts to catch up the BS baseline and N-PPGF baseline. In \figurename \ref{fig_4}\subref{M1} even the $\sigma=0.1$ and the $\epsilon=83428.74$, there is a significant gap between PPGF and DP-SGD. \figurename \ref{fig_5} shows similar trends for the FashionMNIST database. But, we notice that even when $\epsilon = 0.9$, test accuracy yields around 0.8 for all type of model, demonstrating less utility. The ability of VAE generative model is the most significant reason for the performance gap between our method and the source baseline, because there is a certain gap between N-PPGF baseline and the BS baseline. In addition, the performance of DP-SGD is better than PPGF while the $\epsilon$ of DP-SGD is much larger than the PPGF's $\epsilon$. However, lower $\epsilon$ indicates higher indistinguishability, stronger privacy and vice versa. In TABLE \ref{validity of DP-SGD and PPGF}, when $\epsilon$ is close, there is a huge gap between DP-SGD and PPGF. 

\begin{table}[htbp]
\renewcommand{\arraystretch}{1}
% if using array.sty, it might be a good idea to tweak the value of
% \extrarowheight as needed to properly center the text within the cells
\caption{validity of FashionMNIST on several models}
\label{validity of DP-SGD and PPGF}
\centering
\tabcolsep 2pt
\begin{tabular}{|c|c|c|c|c|c|c|c|}
\hline 

\textbf{Architecture}  & \textbf{$\epsilon$} & \textbf{PPGF} & \textbf{$\delta$} & \textbf{$\epsilon$} & \textbf{DP-SGD} \\

\hline
\multirow{3}*{\textbf{ResNet18}} & 0.7 & \textbf{0.7709} & 3.9 & 0.7 & \textbf{0.6542} \\
 & 0.8 & \textbf{0.7805} & 3.4 & 0.81 & \textbf{0.6827} \\
 & 0.9 & \textbf{0.7917} & 3 & 0.92 & \textbf{0.6849} \\
 
\hline
\multirow{3}*{\textbf{ResNet34}} & 0.7 & \textbf{0.7756} & 3.9 & 0.7 & \textbf{0.6118} \\
 & 0.8 & \textbf{0.7788} & 3.4 & 0.81 & \textbf{0.6809} \\
 & 0.9 & \textbf{0.7860} & 3 & 0.92 & \textbf{0.6860} \\
\hline
\multirow{3}*{\textbf{ResNet50}} & 0.7 & \textbf{0.7727} & 3.9 & 0.7 & \textbf{0.6045} \\
 & 0.8 & \textbf{0.7733} & 3.4 & 0.81 & \textbf{0.6411} \\
 & 0.9 & \textbf{0.7881} & 3 & 0.92 & \textbf{0.6615} \\
\hline

\end{tabular}

\end{table}

TABLE \ref{validity of the generated data for CelebA} shows the performance of the synthetic data generated by PPGF with $\epsilon = 0.5$, which is used in the SlimCNN\cite{sharma2019slim} model, as compared to BS baseline, N-PPGF baseline and the DP-SGD baseline in terms of both accuracy for each of the first 12 attributes. We notice that in some attributes like "Blurry", the accuracy of model utilizing synthetic data is almost the same as the accuracy of model using source data. However, in some attributes like "Attractive", there is a huge performance gap between our synthetic data and source data. Even this gap exists between N-PPGF baseline and BS baseline, because the ability of our VAE generative model is limited and some features of raw data can not be reconstructed. In addition, in "Attractive" feature, we also notice that there is also a huge gap between DP-SGD mechanism and source data and we think some feature like "Attractive" is hard to be extracted. Meanwhile, the performance of DP-SGD with $\sigma=0.7$ ($\epsilon=27.06$) is similar with the PPGF ($\epsilon=0.5$). { Lower $\epsilon$ indicates higher indistinguishability, stronger privacy and vice versa. When performance close, PPGF has stronger privacy.}

\begin{table*}[htbp]
% \renewcommand{\arraystretch}{1}
% if using array.sty, it might be a good idea to tweak the value of
% \extrarowheight as needed to properly center the text within the cells
\caption{Performance on SlimCNN for 12 attributes}
\label{validity of the generated data for CelebA}
\centering
\tabcolsep 2pt
\begin{tabular}{|l|c|c|c|c|c|c|c|c|c|c|c|c|}
\hline
\textbf{Features} & \textbf{5\_Clock\_Shadow} & \textbf{Arched\_Eyebrows} & \textbf{Attractive}  \\

\hline
PPGF ($\epsilon=0.5$) & 0.8819 & 0.7408 & 0.5114 \\
DP-SGD ($\sigma=0.7$) & 0.8819 & 0.7415 & 0.5245 \\
N-PPGF & 0.8812 & 0.7416 & 0.5303  \\
BS & 0.9296 & 0.8296 & 0.7841  \\
\hline
\textbf{Features} &  \textbf{Bags\_Under\_Eyes} & \textbf{Bald} & \textbf{Bangs}  \\

\hline
PPGF ($\epsilon=0.5$) & 0.7923 & 0.9793 & 0.8416 \\
DP-SGD ($\sigma=0.7$) & 0.7925 & 0.9793 & 0.8532 \\
N-PPGF & 0.7914 & 0.9793 & 0.8514 \\
BS & 0.8281 & 0.9876 & 0.9474 \\
\hline
\textbf{Features} & \textbf{Big\_Lips} &\textbf{Big\_Nose} & \textbf{Black\_Hair} \\
\hline
PPGF ($\epsilon=0.5$) & 0.8454 & 0.7509 & 0.7534 \\
DP-SGD ($\sigma=0.7$) & 0.8467 & 0.7511 & 0.7914 \\
N-PPGF & 0.8363 & 0.7144 & 0.7717 \\
BS & 0.8487 & 0.8153 & 0.8988 \\
\hline
\textbf{Features} & \textbf{Blond\_Hair} & \textbf{Blurry} & \textbf{Brown\_Hair}  \\
\hline
PPGF ($\epsilon=0.5$) & 0.8443 & 0.9526 & 0.7586 \\
DP-SGD ($\sigma=0.7$) & 0.8461 & 0.9526 & 0.7587 \\
N-PPGF & 0.8462 & 0.9527 & 0.7587 \\
BS & 0.9449 & 0.9595 & 0.8403 \\
\hline
\end{tabular}
\end{table*}

% \begin{table*}[htbp]
% \renewcommand{\arraystretch}{1}
% % if using array.sty, it might be a good idea to tweak the value of
% % \extrarowheight as needed to properly center the text within the cells
% \caption{Performance on SlimCNN for 12 attributes}
% \label{validity of the generated data for CelebA}
% \centering
% \tabcolsep 2pt
% \begin{tabular}{|l|c|c|c|c|c|c|c|c|c|c|c|c|}
% \hline
% \textbf{Features} & \textbf{5\_Clock\_Shadow} & \textbf{Arched\_Eyebrows} & \textbf{Attractive} & \textbf{Bags\_Under\_Eyes} & \textbf{Bald} & \textbf{Bangs}  \\

% \hline
% PPGF ($\epsilon=0.5$) & 0.8819 & 0.7408 & 0.5114 & 0.7923 & 0.9793 & 0.8416 \\
% DP-SGD ($\sigma=0.7$) & 0.8819 & 0.7415 & 0.5245 & 0.7925 & 0.9793 & 0.8532 \\
% N-PPGF & 0.8812 & 0.7416 & 0.5303 & 0.7914 & 0.9793 & 0.8514 \\
% BS & 0.9296 & 0.8296 & 0.7841 & 0.8281 & 0.9876 & 0.9474 \\
% \hline
% \textbf{Features} & \textbf{Big\_Lips} &\textbf{Big\_Nose} & \textbf{Black\_Hair} & \textbf{Blond\_Hair} & \textbf{Blurry} & \textbf{Brown\_Hair}  \\
% \hline
% PPGF ($\epsilon=0.5$) & 0.8454 & 0.7509 & 0.7534 & 0.8443 & 0.9526 & 0.7586 \\
% DP-SGD ($\sigma=0.7$) & 0.8467 & 0.7511 & 0.7914 & 0.8461 & 0.9526 & 0.7587 \\
% N-PPGF & 0.8363 & 0.7144 & 0.7717 & 0.8462 & 0.9527 & 0.7587 \\
% BS & 0.8487 & 0.8153 & 0.8988 & 0.9449 & 0.9595 & 0.8403 \\
% \hline
% \end{tabular}
% \end{table*}

\subsection{Security of PPGF}

We use two membership inference attack methods, namely black-box attack and white-box attack, to attack the target model. For our PPGF, we leverage the generated new data on the three databases to train the target classification model, and utilize the two kinds of membership inference attacks to test. Note that, we choose ResNet-18 model for MNIST, FashionMNIST, CelebA as for the limitation of GPU memory.

\subsubsection{Black-box Attacks}
{We use the black-box attack method in the paper\cite{shokri2017membership, salem2018ml}. We just train one shadow models locally as suggested in \cite{salem2018ml} to learn a two-class model based on the difference between the model's output for member data and non-member data.} Then we use this two-class model to determine whether the data is used for training the target model. The evaluation indicators include the accuracy, precision, recall, fscore, and AUC value of the attack. { Specifically, in this type of attack, there is only 1 shadow model with the same training epochs comparing with the target model and we set a Multi-layer Perceptron(MLP) attack model as paper\cite{salem2018ml} suggested as the two-class attack model. The details of the architecture of the MLP attack model is shown in TABLE \ref{Architecture of Black-box attack model} in Appendix D. As for the training data, when we set the number of training data as $x$, we select the first $x$ data from datasets to train the target model and randomly select $x$ data from raw datasets to train the shadow model. Note that there is no intersection between the training data of the target model and the shadow model. For test data, when we set the number of test data as $x'$, we select randomly select $x'$ data from raw datasets to test the target model and randomly choose $x'$ data from total raw datasets to test the shadow model. All training data and test data of shadow model with corresponding output are used to train our two-class attack model (MLP).} In the end, we use $x$ data from total training data of target model and $x'$ data from total test data of target model to test this black-box attack. The attack results are shown in the TABLE \ref{inference attack on PPGF and BS}. Our framework with $\epsilon = 0.5$ reduces the membership inference accuracy to the target model to be close to $0.5$ and the AUC of the attack model to be close to $0.5$, which means our target model could effectively defend this black-box attack.

\begin{table*}[htbp]
\renewcommand{\arraystretch}{1}
\caption{Membership Inference Attacks on PPGF and BS}
\label{inference attack on PPGF and BS}
\centering
\tabcolsep 0.5pt
\begin{tabular}{|l|c|c|c|c|c|c|c|}
\hline 

\multicolumn{8}{|c|}{\textbf{Black-Box Attack}} \\
\hline
\multicolumn{8}{|c|}{BS(Source Data)} \\
\hline

\textbf{Database} & \textbf{Train size} & \textbf{Test size} & \textbf{Accuracy} & \textbf{Precision} & \textbf{Recall} & \textbf{Fscore} & \textbf{AUC} \\

\hline
% \multirow{6}*{Black-Box}
% & \multirow{3}*{BS(Source Data)}
MNIST & 15000 & 15000 & 0.5149 & 0.5149 & 0.5149 & 0.5149 & 0.5265\\
 \cline{1-8}
FashionMNIST & 15000 & 15000 & 0.6074 & 0.6347 & 0.6074 & 0.5864 & 0.6580\\
 \cline{1-8}
CelebA & 15000 & 15000 & 0.5969 & 0.7718 & 0.5969 & 0.5197 & 0.5896\\
 \cline{1-8}
\hline
\multicolumn{8}{|c|}{PPGF($\epsilon=0.5$)} \\
\hline
\textbf{Database} & \textbf{Train size} & \textbf{Test size} & \textbf{Accuracy} & \textbf{Precision} & \textbf{Recall} & \textbf{Fscore} & \textbf{AUC} \\
\hline
% & \multirow{3}*{PPGF ($\epsilon=0.5$)}
MNIST & 15000 & 15000 &\textbf{0.4966} & 0.4965 & 0.4966 & 0.4934 & \textbf{0.4971}\\
 \cline{1-8}
FashionMNIST & 15000 & 15000 & \textbf{0.5025} & 0.5025 & 0.5025 & 0.5020 & \textbf{0.5045}\\
 \cline{1-8}
CelebA & 15000 & 15000 & \textbf{0.4967} & 0.4951 & 0.4967 & 0.4522 & \textbf{0.4971}\\
 \cline{1-8}
\hline
% \hline
% \multirow{6}*{White-Box}
% & \multirow{3}*{BS(Source Data)}
\multicolumn{8}{|c|}{\textbf{White-Box Attack}} \\
\hline
\multicolumn{8}{|c|}{BS(Source Data)} \\
\hline

\textbf{Database} & \textbf{Train size} & \textbf{Test size} & \textbf{Accuracy} & \textbf{Precision} & \textbf{Recall} & \textbf{Fscore} & \textbf{AUC} \\

\hline

MNIST & 2000 & 2000 & 0.5415 & 0.5544 & 0.5415 & 0.5126 & 0.5699\\
  \cline{1-8}
FashionMNIST & 2000 & 2000 & 0.5815 & 0.7295 & 0.5815 & 0.5011 & 0.6072 \\
  \cline{1-8}
CelebA & 2000 & 2000 & 0.5425 & 0.6197 & 0.5425 & 0.4546 & 0.5639 \\
  \cline{1-8}
\hline
% & \multirow{3}*{PPGF ($\epsilon=0.5$)}
\multicolumn{8}{|c|}{PPGF($\epsilon=0.5$)} \\
\hline
\textbf{Database} & \textbf{Train size} & \textbf{Test size} & \textbf{Accuracy} & \textbf{Precision} & \textbf{Recall} & \textbf{Fscore} & \textbf{AUC} \\
\hline
MNIST & 2000 & 2000 & \textbf{0.5015} & 0.5319 & 0.5015 & 0.3452 & \textbf{0.4986}\\
  \cline{1-8}
FashionMNIST & 2000 & 2000 & \textbf{0.5085} & 0.5115 & 0.5085 & 0.4745 & \textbf{0.5176} \\
  \cline{1-8}
CelebA & 2000 & 2000 & \textbf{0.5020} & 0.5459 & 0.5020 & 0.3455 & \textbf{0.4944} \\
  \cline{1-8}
\hline
\end{tabular}
\end{table*}

\subsubsection{White-box Attacks}
We utilize the white-box attack method from the paper \cite{nasr2019comprehensive}. As is shown in this paper, the white-box attack is more effective than the black-box one and considered to be the most effective attack. The adversary controls the target model and could test target models locally, learning a two-class model based on the difference between the model's layer output, loss values, label values, and layer gradients for member data and non-member data. In this part, we use this two-class model in \cite{nasr2019comprehensive} to determine whether the data is used for training the target model. The evaluation indicators include the accuracy, precision, recall, fscore and AUC value of the attack. Specifically, we only select the gradients of last layer, the model's output, loss values and label values to train this two-class model. { We use 60000 training data to train target model. And we set training size as $x$ and test size as $x'$. We randomly select $x$ data from all raw training data and $x'$ data from all raw test data to test this white-box attack.} As shown in the TABLE \ref{inference attack on PPGF and BS}, our framework with $\epsilon = 0.5$ reduces the inference accuracy to the target model to be close to $0.5$ and the AUC of the attack model to be close to $0.5$, which means our target model could prevent this white-box attack.

\section{Related Work}\label{section6}

\subsection{Membership Inference Attack} 
The first paper about the membership inference attack was published by Shokri \emph{et al}. \cite{shokri2017membership} under the assumption of the black-box model. The shadow model was used to train a machine learning model similar to the target model, and the shadow model was used to learn differences of confidence vector about members and non-members. Later, Nasr \emph{et al.}\cite{nasr2019comprehensive} first proposed the membership inference attack under the white-box model and federated learning model. They classified between members and non-members by distinguishing model data such as the gradient changes of each layer or the output of each layer. Song \emph{et al.}\cite{song2019privacy} mainly discussed the practical threat of machine learning in the article. The defensive measures against adversarial samples were to generate a robust machine learning model. But, this robust machine learning model was more sensitive to membership inference attacks. The article \cite{song2021systematic} set different thresholds, which was used to determine whether the prediction confidence belongs to the member, for a different class label to improve the accuracy of the attack. At the same time, the redefined prediction entropy was used to strengthen the membership inference attack. The paper \cite{yu2020membership} designed new membership inference algorithms against machine learning models and achieved significantly higher inference accuracy when the augmented data was also used in training but the augmented mechanism was known to the adversary. However, the malicious attacker could only utilize the original image or randomly chosen transformations which yielded a significantly lower inference success rate. There are also some membership inference attacks where the adversary's advantages are limited. Hui \emph{et al.}\cite{hui2021practical} proposed an membership inference attack when the adversary couldn't collect enough sample with output probabilities and labels as either members or non-members and Choquette \emph{et al.}\cite{choquette2021label} introduced label-only membership attacks with the adversary only got access to models' predicted labels. In addition, there was an new membership inference attack \cite{zhang2021membership} to recommander systems where the adversary could only observe the ordered recommended items.

\subsection{Protection Mechanism Against Membership Inference Attack}
Membership inference attack defense mechanisms have also developed for many years. As mentioned in the article\cite{salem2018ml}, they used the dropout mechanism and model stacking mechanism to avoid over-fitting of the target model and prevent membership inference attacks. Song \emph{et al.} \cite{nasr2018machine} introduced adversarial training in the training process, which was abstracted into a maximum and minimum problem. Minimizes the classification loss and maximizes the profit of the inference attack to find the best defensive model against the attack model. Jia \emph{et al.}\cite{jia2019memguard} studied to fuzzy the confidence score vector to evade the membership classification of the attack model. Wang \emph{et al.}\cite{wang2020against} reduced model storage and computational operation to defend membership inference attacks. Yin \emph{et al.} \cite{yin2021defending} set definitions for utility and privacy of target classifier, and formulated the design goal of the defense method as an optimization problem and solved it. 

\textbf{Differential Privacy:} For the schemes that apply differential privacy technology to protect privacy, some schemes\cite{bassily2014private, wang2017differentially, iyengar2019towards, jayaraman2019evaluating, nasr2021adversary} added noise to the objective function to protect privacy, while others \cite{abadi2016deep, yu2019differentially, mcmahan2018general} called DP-SGD added noise to the gradient during training. Jayaraman \emph{et al.} \cite{jayaraman2019evaluating} added noise in the objective function, gradient vectors, and output vectors to analyze trade-offs between privacy and utility. Nasr \emph{et al.} instantiated the hypothetical adversary to analyze the differential private machine learning especially DP-SGD. There are also some schemes\cite{hayes2019logan, chen2019gan, chen2021gan, liu2019performing, hu2021membership} that set the goal of protection as a generative model. Hayes\emph{et al.} \cite{hayes2019logan} first applied the membership inference attack on the generative model and utilized differential privacy during training stage to protect the generative model. Chen \emph{et al.} \cite{chen2019gan} Systematically analyzed the membership inference attack on the generative model and made a detailed classification. However, we treat generative model VAE as a building component to protect the downstream task model in this work. 

The existing defense schemes basically add noise in the model training process or directly confuse the output results of the model. It is rarely considered to protect the privacy of the source data by processing the raw data. 

\subsection{Data Generation Mechanism}
Some data generation mechanisms\cite{xie2018differentially, torkzadehmahani2019dp, zhang2018differentially, chen2020gs} satisfying differential privacy have been proposed, all of these works considered that leverage differential noise mechanism in the training process as previously stated and the membership inference attacks were not included. Xie \emph{et al.} \cite{xie2018differentially} added noise on the gradient of the Wasserstein distance with respect to the training data during the training procedure. The DP-CGAN \cite{torkzadehmahani2019dp} made the training process private by injecting random Gaussian noise into the optimization process of the discriminator network. DP-GAN \cite{zhang2018differentially} was built upon the improved WGAN framework and enforced DP by injecting random noise in updating the discriminator. Chen \emph{et al.} \cite{chen2020gs} proposed gradient-sanitized Wasserstein GAN replacing the typical training procedure with the DP-SGD. However, we consider that generating synthetic data to train the target model for source data protection and don't affect the training process of the generative model. On the one hand, we don't consider applying the differential privacy mechanism during the training stage because of the model convergence, privacy budget costs in each iteration, and schemes extension. On the other hand, their purpose is to construct generative models satisfying differential privacy to generate lots of new data, while this paper focuses on protecting the target model from membership inference attacks by transforming the original data into similar data that satisfy differential privacy.

\section{Conclusion and Future Work}\label{section7}
In this paper, we have proposed a privacy-preserving system framework for preventing source data leakage to generate synthetic data satisfying differential privacy without much loss. This is achieved via a well-trained VAE generative model and an extended differential privacy mechanism. We first gain a VAE model through our source data. Next, we can use a VAE encoder to gain data feature from the data source and add some noise which satisfies extended differential privacy. Then we put the result into the VAE decoder to get new data. Finally, we can use synthetic data to train the target machine learning model. Therefore, our proposed framework can apply in any downstream machine learning environment without any change to the target model. Extensive experiment results also show the effectiveness and generalization ability of our framework.
Recall that the ability of the generative model to extract data feature limits the validity of the generated data. Thus, an interesting future work is providing an application programming interface to the more complex and effective generative model in our framework and any latent generative model can apply in this framework. Another direction is to discover the fine-grained noise mechanism which can make the synthetic data more realistic in a specific feature.

% \subsection*{Acknowledgements}
% This work was supported by the Foundation for Innovative Research Groups of the National Natural Science Foundation of China (Grant No. 62121001), the National Natural Science Foundation of China (Grant Nos. 61902290, 62072352, 61872283), Key Research and Development Program of Shaanxi (Grant No. 2020ZDLGY09-06), Scientific Research Program Funded by Shaanxi Provincial Education Department (Grant No. 20JY016).

\appendix

% \addcontentsline{toc}{section}{Appendix}
% \section{My Appendix}
% Appendix sections are coded under \verb+\appendix+.
\setcounter{equation}{0}
\renewcommand{\theequation}{A.\arabic{equation}}
\section{Coefficient of Probability Density Function}
The derivation of derivates the coefficient $\lambda_{\epsilon,k}$ from the pdf $D_{\epsilon,k}(x_0)(x)$ is as follows. We convert the Cartesian coordiantes to the hyper-spherical coordinate system and set the $x_0$ at the origin of the hyper-spherical coordinate system without loss generality.
\begin{align}
\label{equation10}
\nonumber \int_{R^k}D_{\epsilon,k}(x_0)(x)dx &= \int_{0}^{\infty}\int_{\sum} \lambda_{\epsilon,k} e^{-\epsilon\cdot r} dsdr\\
\nonumber &= \int_{0}^{\infty} \lambda_{\epsilon,k} \cdot e^{-\epsilon\cdot r} \cdot \int_{\sum} dsdr  \\
\nonumber &= \int_{0}^{\infty} \lambda_{\epsilon,k} \cdot e^{-\epsilon\cdot r} \frac{2\pi^{k/2}}{\Gamma(\frac{k}{2})} \cdot r^{k-1} dr \\
\nonumber &= \lambda_{\epsilon,k} \cdot \frac{2\pi^{k/2}}{\Gamma(\frac{k}{2})} \cdot\int_{0}^{\infty} e^{-\epsilon\cdot r} \cdot r^{k-1} dr \\
 &= \lambda_{\epsilon,k} \cdot \frac{2\pi^{k/2}}{\Gamma(\frac{k}{2})} \cdot \frac{\Gamma(k)}{\epsilon^k} = 1
\end{align}

From the (\ref{equation10}) and $\Gamma(n+1) = n!$, we gain: 
\begin{equation}
\lambda_{\epsilon,k} = \frac{1}{2}\cdot(\frac{\epsilon}{\sqrt{\pi}})^k\cdot\frac{(\frac{k}{2}-1)!}{(k-1)!}
\end{equation}

\section{Proof of the Marginal Distribution}
We could gain the marginal distribution function of the pdf $D_{\epsilon,k}(x_0)(x)$ as follows.
\begin{align}
\nonumber \int_{\sum}D_{\epsilon,k}(x_0)(x)dx &= \int_{\sum} \lambda_{\epsilon,k} e^{-\epsilon\cdot r} ds\\
\nonumber &= \lambda_{\epsilon,k} \cdot e^{-\epsilon\cdot r} \cdot \int_{\sum} ds \\
\nonumber &=  \lambda_{\epsilon,k} \cdot e^{-\epsilon\cdot r} \cdot \frac{2\pi^{k/2}}{\Gamma(\frac{k}{2})} \cdot r^{k-1} \\
&= \frac{\epsilon^{k}}{\Gamma(k)} \cdot  e^{-\epsilon\cdot r} \cdot r^{k-1}
\end{align}

The pdf of $\Gamma$-distribution is as follows:
\begin{equation}
f(x,\beta,\alpha)=\frac{\beta^{\alpha}}{\Gamma(\alpha)} \cdot e^{-\beta\cdot x} \cdot x^{\alpha-1}
\end{equation}

we could replace $\alpha$ with k and replace $\beta$ with $\epsilon$. Obliviously, the marginal distribution satisfies $\Gamma$-distribution. 
% \section{}

\section{Architecture of Variational Autoencoder Models}
\setcounter{table}{0}
\renewcommand\thetable{\Alph{section}\arabic{table}} 
\begin{table}[H]
\renewcommand{\arraystretch}{1}
\caption{Architecture of VAE model for MNIST and FashionMNIST}
\label{Architecture of VAE model for MNIST and FashionMNIST}
\centering
\begin{tabular}{|l|l|l|c|c}
\hline 
\textbf{Name} & \textbf{Layer Type} & \textbf{Details}  \\
\hline
% \cline{1-3}
\multirow{3}*{Encoder Component} & \multirow{2}*{Fully Connected Layer} & Sizes: 768 $\rightarrow$ 400\\
% \cline{2-3}
& & Activation: ReLU\\
\cline{2-3}
& Fully Connected Layer & Sizes: 400 $\rightarrow$ 20\\
\hline
 \multirow{3}*{Decoder Component} & \multirow{2}*{Fully Connected Layer} & Sizes: 20 $\rightarrow$ 400\\
% \cline{2-3}
& & Activation: ReLU\\
\cline{2-3}
& Fully Connected Layer & Sizes: 400 $\rightarrow$ 768\\
\hline
\end{tabular}
\end{table}

\begin{table}[H]
\renewcommand{\arraystretch}{1}
\caption{Architecture of VAE model for CelebA}
\label{Architecture of VAE model for CelebA}
\centering
\begin{tabular}{|l|l|l|c|c}
\hline 
\multicolumn{2}{|c|}{\textbf{Encoder Component}} \\
\hline 
\textbf{Layer Type} & \textbf{Details}  \\
\hline
% \cline{1-3}
% \multirow{26}*{Encoder Component} & \multirow{5}*{Convolutional Layer} & Kernels: 128\\
\multirow{5}*{5 Convolutional Layers} & Kernels: 128, 256, 512, 1024, 2048\\
% \cline{2-3}
 & Kernel size$: 4 \times 4$\\
% \cline{2-3}
 & Stride: 2\\
 & Batch Normalization\\
 & Activation: ReLU \\
% \hline
\cline{1-2}
  Fully Connected Layer & Size: 32768 $\rightarrow$ 512\\
\hline
\multicolumn{2}{|c|}{\textbf{Decoder Component}} \\
\hline
\textbf{Layer Type} & \textbf{Details}  \\
\hline
  \multirow{2}*{Fully Connected Layer} & Size: 512 $\rightarrow$ 32768\\
 & Activation: LeakyReLU \\
% \cline{1-2}
\cline{1-2}
 \multirow{5}*{6 Convolutional Layers}  & Kernels: 2048, 1024, 512, 256, 128, 3\\
% \cline{2-3}
 & Kernel size$: 4 \times 4$\\
% \cline{2-3}
 & Stride: 2\\
 & Batch Normalization \\
 & Activation: LeakyReLU\\
% \hline
\cline{1-2}
\hline
\end{tabular}
\end{table}

\section{Architecture of Black-box Attack Model}
\setcounter{table}{0}
\renewcommand\thetable{\Alph{section}\arabic{table}} 
\begin{table}[H]
\renewcommand{\arraystretch}{1}
\caption{Architecture of Black-box attack model}
\label{Architecture of Black-box attack model}
\centering
\begin{tabular}{|l|l|l|c|c}
\hline 

\textbf{Name} & \textbf{Layer Type} & \textbf{Details}  \\
% \cline{1-3}
\hline
\multirow{3}*{MLP Component} & \multirow{2}*{Fully Connected Layer} & Sizes: len(output) $\rightarrow$ 128\\
% \cline{2-3}
& & Activation: ReLU\\
\cline{2-3}
& Fully Connected Layer & Sizes: 128 $\rightarrow$ 2\\
\hline
%  \multirow{3}*{Decoder Component} & \multirow{2}*{Fully Connected Layer} & Sizes: 20 $\rightarrow$ 400\\
% % \cline{2-3}
% & & Activation: ReLU\\
% \cline{2-3}
% & Fully Connected Layer & Sizes: 400 $\rightarrow$ 768\\
% \hline

\end{tabular}

\end{table}

\bibliography{informationscience}

\end{document}